\documentclass[useAMS,usenatbib,usegraphicx]{mn2e}

\def\ngc{NGC\,6231}
\def\hda{HD\,152248}
\def\hdb{HD\,93403}
\def\hdc{HD\,159176}
\def\hdd{HD\,93205}
\def\sgr{9\,Sgr}
\def\etacar{$\eta$\,Car}
\def\g2vel{$\gamma^2$\,Vel}

\def\sco{Sco\,OB\,1}

\def\chin{$\chi^2_{\nu}$}

\def\l{$\lambda$\,}
\def\halpha{H\,$\alpha$}

\def\heb{He\,{\sc ii}}

\def\kms{km\, s$^{-1}$}
\def\cnts{cnt\ s$^{-1}$}
\def\ergs{erg\ s$^{-1}$}
\def\rsol{R$_{\odot}$}
\def\msol{M$_{\odot}$}

\def\xmmnew{{\sc XMM}\emph{-Newton}}

\def\rosat{\emph{{\sc ROSAT}}}

\def\einstein{\emph{{\sc EINSTEIN}}}
\def\pspc{{\sc PSPC}}
\def\rosatpspc{{\sc ROSAT PSPC}}
\def\xmm{{\sc XMM}\emph{-Newton}}
\def\xmmepic{{\sc XMM EPIC}}
\def\epic{{\sc EPIC}}
\def\mos{{\sc MOS}}
\def\pn{pn}
\def\rgs{{\sc RGS}}
\def\epicmos{{\sc EPIC MOS}}
\def\epicpn{{\sc EPIC} pn}
\def\xspec{{\sc xspec}}
\def\mekal{{\sc mekal}}

\title[A phase-resolved \xmmnew\ Campaign on the Colliding Wind Binary HD 152248]{A  phase-resolved \xmmnew\ Campaign on \\the Colliding Wind Binary HD 152248\thanks{Based on observations collected with \xmmnew, an ESA Science Mission with instruments and contributions directly funded by ESA Member States and the USA (NASA).}}

\author[H. Sana, I.R. Stevens, E. Gosset, G. Rauw and J.-M. Vreux]{
H. Sana$^{1}$\thanks{E-mail:
sana@astro.ulg.ac.be}\thanks{Research Fellow FNRS (Belgium)},
I.R. Stevens$^{2}$,
E. Gosset$^{1}$\thanks{Research Associate FNRS (Belgium)},
 G. Rauw$^{1}$\footnotemark[4] and 
J.-M. Vreux$^{1}$ \\
$^{1}$Institut d'Astrophysique et de G\'eophysique, Universit\'e de Li\`ege, Bat. B5c,  All\'ee du VI Ao\^ut 17, B-4000 Li\`ege, Belgium\\
$^{2}$School of Physics \& Astronomy, University of Birmingham, Edgbaston, Birmingham B15 2TT, United Kingdom}
\begin{document}

\date{Accepted 1988 December 15. Received 1988 December 14; in original form 1988 October 11}

\pagerange{\pageref{firstpage}--\pageref{lastpage}} \pubyear{2002}

\maketitle

\label{firstpage}

\begin{abstract}
We report the first results of an \xmmnew\ monitoring campaign of the \ngc\ open cluster in the \sco\ association. This first paper focuses on the massive colliding wind binary \hda, which is the brightest X-ray source of the cluster. The campaign, with a total duration of 180\,ksec, was split into six separate observations, following the orbital motion of \hda. The X-ray flux from this system presents a clear, asymmetric modulation with the phase and ranges from $0.73$ to $1.18\,10^{-12}$ \ergs\,cm$^{-2}$ in the [0.5-10.0\,keV] energy band. The maximum of the emission is reached slightly after apastron. The \epic\ spectra are quite soft and peak around 0.8-0.9\,keV. We characterize their shape using several combinations of \mekal\ models and power-law spectra and we detect significant spectral variability in the [0.5-2.5\,keV] energy band. 

We also perform 2-D hydrodynamical simulations using different sets of parameters that closely reproduce the physical and orbital configuration of the \hda\ system at the time of the six \xmm\ pointings. This allows a direct confrontation of the model predictions with the constraints deduced from the X-ray observations of the system. We show that the observed variation of the flux can be explained by a variation of the X-ray emission from the colliding wind zone, diluted by the softer X-ray contribution of the two O-type stars of the system. Our simulations also reveal that the interaction region of \hda\ should be highly unstable, giving rise to shells of dense gas that are separated by low density regions. 

Finally, we perform a search for short-term variability in the light curves of the system and we show that trends are present within several of the 30\,ksec exposures of our campaign. Further, most of these trends are in good agreement with the orbital motion and provide a direct constraint on the first order derivative of the flux. In the same context, we also search for long-range correlations in the X-ray data of the system, but we only marginally detect them in the high energy tail of the signal.
\end{abstract}


\section{Introduction}
\label{sect: intro}

Early-type stars of spectral type O and their evolved descendants, the Wolf-Rayet stars, are among the most luminous and hottest objects of the Milky Way. Characterized by their strong and powerful winds, and their huge mass loss rates, they are also  known to be X-ray emitters since the late 70's and the advent of the X-ray observatories equipped with Wolter focusing mirrors. Though with some dispersion, the X-ray emission from single O and early B stars approximately scales with their bolometric luminosity: $L_\mathrm{X} \sim 10^{-7} L_\mathrm{bol}$ \citep{Berg97} and displays a soft thermal spectrum ($kT\sim$\ 0.2-1.0\,keV). The exact mechanism of this X-ray emission, however, is still not fully understood but it is generally thought that, in the lower layers of the winds,  the plasma is heated in shocks that arise from small-scale wind structures that grow out from line-driven wind instabilities \citep*{Feld97,DO03}. \\

Beyond this general scheme,  it is now established that massive binaries often display an additional X-ray luminosity compared to single stars of the same spectral types and luminosity classes \citep{CG91}. The extra X-ray component is usually attributed to emission from a hot plasma that results from the hydrodynamical collision of the winds of the two stars. The \emph{standard} collision model predicts the interaction region to be limited by two curved shock-surfaces between which the gas is heated up to temperatures of a few $10^8 K$.
From the observational point of view, this phenomenon can manifest itself i) as a modulation of the X-ray emission due to the variation of the line-of-sight opacity with the orbital motion,  ii) in an eccentric binary, as a variation of the intrinsic emission due to a variation of the separation between the two components, as the densities of the colliding material and, in close binaries, the pre-shock wind velocities  depend on the distance to the originating  star. Though the X-ray domain is probably the most adequate to study the wind interaction region in close binary systems, the observational signature of the collision is not restricted to high energies. From the UV and optical domains for example, the interaction may be traced in emission lines that are produced by recombination of the cooling gas that escapes the interaction region after the shocks \citep*[see e.g.][]{SRG01,Rauw02}. \\

In addition to the improvements in observational capabilities, another tool for probing the physics of wind-wind collisions has undergone important development in the past 10 to 15 years. Namely Computational Fluid Dynamic (CFD) applied to the hydrodynamical problem of wind-wind collision holds out promise of insight into the physics of the phenomenon \citep*{SBP92,PS97,HSP03,Deb03}. 
For example, from the work of \citet[ SBP92 hereafter]{SBP92}, it is obvious that the shocks and the interaction zone properties are strongly dependent on the characteristics of the fluids upstream the shock. In a first step to classify the shocks, SBP92 introduced a parameter $\chi=t_\mathrm{cool} / t_\mathrm{flow}$ that represents the ratio of the characteristic cooling time of the fluid  downstream the shock ($t_\mathrm{cool}$) to the characteristic flow time ($t_\mathrm{flow}$). Based on this simple criterion, two extreme behaviours might be distinguished. On one hand the gas escapes the interaction region with almost no significant cooling ($\chi>>1$) and the phenomenon is mostly adiabatic. On the other hand, whenever $\chi \la 1$, the radiative cooling then plays a crucial role. In the adiabatic case, the escaping flow is expected to be relatively smooth and the region in between the shocks is quite large. When the flow is dominated by radiative cooling, however, the shock region collapses and the steady state is disrupted by Kelvin-Helmholtz and thermal instabilities (SBP92). \\

However until recently very few X-ray observations of O+O Colliding Wind Binaries (CWBs) had a sufficient quality (counts, time resolution, ...) to allow a detailed comparison with predictions from CFD calculations. 
%Furthermore, most of the studies using the previous generation of X-ray satellites (\asca, \rosat, ...) focused on WR+O binaries (e.g. \g2vel, \citet{g2vel}, ...) which are characterized by a wind momentum balance that is located far from the unity.  
With the advent of the \xmmnew\ and \emph{Chandra} X-ray observatories, the hot star community was offered for the first time the possibility of acquiring high quality data that would provide a strong test of the latest models. This should therefore noticeably contribute to improving our understanding of X-ray emission from early-type stars, and more specifically of the winds of these particular objects. However, despite their interest, still very few O+O binaries have been the target of such an X-ray monitoring to detect the expected phase-locked behaviour of the emission. \\

\hdb\ is an O5.5I + O7V binary with an orbital period close to 15.1 days and an eccentricity of 0.234 \citep{Rauw00}. An \xmm\ campaign revealed a phase locked behaviour of the X-ray flux that is mostly consistent with a $1/D_{\mathrm{sep}}$ dependence, where $D_{\mathrm{sep}}$ is the separation between the two stars, as expected for an adiabatic wind-wind interaction \citep{Rauw02}. 

A similar behaviour is observed in the case of \hdd, an eccentric system ($e=0.37$, $P_{\mathrm{orb}}=6.08$\,days) consisting of an O3\,V primary and an O8\,V secondary, in the Tr\,16 cluster near \etacar. \citet{Antokhin} analyze five \xmm\ observations of the region around \etacar. They report a clear phase-locked modulation of the flux, with maximum emission occurring at periastron and minimum around apastron.

In the case of the O7\,V\,+\,O7\,V binary \hdc\ ($P_{\mathrm{orb}} = 3.367$\,days), \citet{Deb03} obtained a single observation with \xmm. The system was found to display a rather modest X-ray overluminosity attributed to a wind collision. The authors further find that only part of the wind kinetic power is actually emitted in the X-ray domain as a result of the wind interaction.\\

\begin{table}
\caption{ 
\label{tab: orbit} 
Orbital and physical parameters of the \hda\ binary. The usual notations have been adopted. $T_0$ is the time of periastron passage and is adopted as phase $\phi=0.0$. Most of the quoted values are from Paper I. Whenever this is not the case,  reference to the original paper is provided.
}
\begin{center}
\begin{tabular}{l r c l }
\hline
$P_\mathrm{orb}$ (days) & 5.816\,032 & $\pm$ & 0.000\,058 \\
$e$                   &    0.127 & $\pm$ & 0.007 \\
$\omega$ (\degr)      &     84.8 & $\pm$ & 4.7   \\    
$i$ (\degr)           & 67.2$^a$ &             \\
$T_0$(HJD             & 2003.879 & $\pm$ & 0.072 \\
\hspace*{2mm} $-$2\,450\,000) &  &             \\
$\gamma_1$ (\kms)     &  $-30.3$ & $\pm$ & 1.5   \\
$K_1$ (\kms)          &    216.0 & $\pm$ & 1.5   \\
$a_1\sin i$ (\rsol)   &    24.59 & $\pm$ & 0.17  \\ 
$\gamma_2$ (\kms)     &  $-28.7$ & $\pm$ & 4.3   \\
$K_2$ (\kms)          &    213.7 & $\pm$ & 5.2   \\
$a_2\sin i$ (\rsol)   &    24.35 & $\pm$ & 0.62  \\
 \\		        
$q\ (=M_1/M_2)$       &    0.990 & $\pm$ & 0.023  \\
$M_1\sin^3 i$ (\msol) &    23.19 & $\pm$ & 1.19  \\
$M_2\sin^3 i$ (\msol) &    23.44 & $\pm$ & 0.73  \\		
R$_1$ (\rsol)         &     15.4 & $\pm$ & 3.5   \\ 
R$_2$ (\rsol)         &     14.9 & $\pm$ & 3.5   \\
\\
v$_\infty$ (\kms)     & 2420$^{b,*}$       \\
log $\dot{M}_1$ (\msol\,yr$^{-1}$)& $-5.53$          \\
log $\dot{M}_2$ (\msol\,yr$^{-1}$)& $-5.48$          \\
d (pc)                & 1757 & $\pm$ &370 $^{c}$ \\
\hline

\end{tabular}
\begin{tabular}{l}
a. \citet{MHL01} \\
b. \citet{HSHP97} \\
c. \citet*{RCB97} \\
$*$. see Sect. \ref{ssect: hcode}\\
\end{tabular}
\end{center}
\end{table}

In this context, we undertook a phase-resolved \xmm\ campaign targeting the early-type binary \hda. This system lies at the centre of the young open cluster \ngc\ in the core of the \sco\ association.
It is a close SB2 eclipsing binary consisting of two almost identical O-type giants with a period close to 5.8 days.
The orbital and physical properties of the system were established by \citet[ Paper I]{SRG01} from medium and high resolution optical spectroscopy. In Paper I, we showed that \hda\ is an O7.5(f)\,III\,+\,O7(f)\,III system with the O7.5\,III primary being slightly less massive than the secondary. The convention that refers to the less massive component as the primary star historically results from the light curve of the system that displays a deeper minimum during the occultation of the O7.5 star. Table \ref{tab: orbit} provides the physical and orbital parameters of the system. 
Though \hda\ could not be resolved with the \einstein\ observatory due to the limited spatial resolution of the satellite and the crowded nature of the field, \hda\ was observed by the R\"ontgen Satellite (\rosat) and clear variations of the flux with the orbital phase could be identified \citep{Corc96}.
Paper I also presents strong clues  that \hda\ harbours a colliding wind phenomenon that could be tracked in the optical domain through the  phase-locked profile variations of the \heb\,\l4686 and \halpha\  emissions. \\

 It therefore appeared extremely promising to compare the latest X-ray observations of this object with the predictions of recent hydrodynamical models. Indeed the quality of the \hda\ \xmm\ data combined with the very good knowledge of the system and the high level of constraints on its physical and geometrical parameters (including the usually rather tricky determination of the inclination) provide the opportunity i) to perform numerical simulations corresponding to a well constrained configuration, and consequently very representative of the \hda\ geometry, ii) to provide predictions that are suitable for a direct confrontation with observational results, iii) to test the state-of-the-art CWB models, iv) to gain insight into the physics of wind-wind collisions and v) to uncover new routes to improve hydrodynamical simulations of close binary systems. In this paper, we thus discuss the X-ray emission from the \hda\ system both from the observational point of view and by means of hydrodynamical calculations of wind-wind collisions. The analysis of the other X-ray sources of the field will be addressed in subsequent papers. We further focus on the \epic\ observations of the system and on a detailed comparison with corresponding numerical simulations. We do not tackle here the analysis of the \rgs\ data. Indeed, as the field around \hda\ is relatively crowded,  the \rgs\ spectra are contaminated by rather bright neighbouring sources and a simultaneous analysis of the \epic\ data of the contaminants as well as specific techniques to estimate the level of the contamination are required. This is beyond the scope of this study and will be deferred to a future work. \\

The present paper is organised as follows. The next section presents a description of the \xmm\ campaign and the data reduction. Sect. \ref{sect: epic_sciences} is focused on \xmmepic\ spectra analysis and, in Sect. \ref{sect: simul}, we describe the performed hydrodynamical simulations. We also provide a detailed comparison  between observation and simulation results. The next section (Sect. \ref{sect: st_var}) investigates the short-term variability of the observed X-ray emission. Finally,  Sect. \ref{sect: concl} presents a summary of the main results of this paper as well as the conclusions of the present work.\\

\begin{figure}
\begin{center}
% \rotatebox{-90}{
% \includegraphics[height=8cm]{fig1.ps}}
\caption{Scale sketch, in the orbital plane of the system, representing the configuration of the \hda\ binary at the time of the six \xmmnew\ pointings. The primary star is in dark grey while the secondary is represented in light grey. Arrows at lefthand indicate the projection, onto the orbital plane, of the line of sight of the satellite towards the system. 
\label{fig: config}  }
\end{center}
\end{figure}

\begin{table*}
\caption{ 
\label{tab: journal} 
Journal of the \xmm\ observations of \hda. Columns 2 and 3 give the spacecraft revolution number and the observation ID. The mean Julian Day (JD) is reported in Col. 4. The next three columns list the performed exposure times for the  \epicmos, \epicpn\ and \rgs\ instruments. The last column provides the orbital phase of \hda\ for each \xmm\ observation at mid-exposure, according to the ephemerides given in Table \ref{tab: orbit}. The  quoted {\it uncertainties} stand for the phase intervals that correspond to the durations of each observation.
}
\begin{center}
\begin{tabular}{c c c c c c c c c}
\hline
Obs. \# & Sat. & Exposure & Mean JD & \multicolumn{3}{c}{Performed duration (ksec)} & \hda's \\
       & Rev. &  ID      &JD$-2\,450\,000$& \mos & \pn & \rgs & phase \\ 
\hline
1 & 319 & 0109490101 & 2158.214 & 33.3 & 30.7 & 33.8 & 0.54 $\pm$ 0.03 \\
2 & 319 & 0109490201 & 2158.931 & 22.1 & 20.2 & 22.4 & 0.66 $\pm$ 0.02 \\
3 & 320 & 0109490301 & 2159.796 & 34.4 & 31.8 & 35.0 & 0.81 $\pm$ 0.03 \\
4 & 320 & 0109490401 & 2160.925 & 31.4 & 29.1 & 31.5 & 0.00 $\pm$ 0.03 \\
5 & 321 & 0109490501 & 2161.774 & 31.1 & 28.5 & 31.6 & 0.15 $\pm$ 0.03 \\
6 & 321 & 0109490601 & 2162.726 & 32.9 & 30.3 & 33.5 & 0.31 $\pm$ 0.03 \\
\hline
\end{tabular}
\end{center}
\end{table*}

\section{Observations and Data Reduction }\label{sect: reduc}
\subsection{The {\sc XMM}{\it -Newton} observing campaign }\label{ssect: campaign}
The X-ray observational campaign was designed in the framework of the \xmm\ guaranteed time of observation of the {\sc XMM-OM} consortium. It initially consisted of six 30\,ksec pointings towards the young open cluster \ngc\ that shelters the \hda\ binary. These six observations were spread over the 5.816 day period of the system (see Fig. \ref{fig: config}) to monitor the expected phase-locked behaviour of the emission. In September 2001, the \xmmnew\ satellite successfully performed the six observations within satellite revolutions 319 to 321, i.e. within a single orbital cycle of \hda. Two of the six data sets were unfortunately affected by soft proton flares which effectively reduced the observing time by about one third. Table \ref{tab: journal} gives the journal of the X-ray campaign dedicated to the cluster \ngc. All three \epic\ instruments were operated in the Full Frame mode together with the {\sc thick} filter to reject optical light. The \rgs\ spectrographs were run in the Standard Spectroscopic mode. Due to the brightness of the objects in the field of view (FOV) of \ngc, the Optical Monitor was switched off throughout the campaign.\\

\begin{figure}
\begin{center}
% \rotatebox{-90}{
% \includegraphics[height=6cm]{fig2a.ps}}
% \rotatebox{-90}{
% \includegraphics[height=6cm]{fig2b.ps}}
\caption{
\label{fig: xmmfov} \mos1 (\emph{upper panel}) and \pn\ (\emph{lower panel}) images of the central part of the \ngc\ cluster in the energy band [0.5-2.0\,keV]. Adopted extraction regions for \hda\ and for the background are also indicated. }
\end{center}
\end{figure}

\subsection{\xmmepic\ data reduction }\label{ssect: epic_data_red}
The \epic\ Observation Data Files (ODFs) were processed using the XMM-Science Analysis System (SAS) v5.2 implemented on our computers in Li\`ege. We applied the {\it emproc} and {\it epproc} pipeline chains respectively to the \mos\ and \pn\ raw data to generate proper event list files. No indication of pile-up was found in the data. We then only considered  events with patterns 0-12 (resp. 0) for \mos\ (resp. \pn) instruments and we applied the filtering criteria XMMEA\_EM and XMMEA\_EP as recommended by the Science Operation Centre (SOC) technical note XMM-PS-TN-43\,v2.0. For each pointing, we rejected periods affected by soft proton flares. For this purpose, we built light curves at energies above 10\,keV\footnote{Expressed in Pulse Invariant (PI) channel numbers and considering that 1 PI channel approximately corresponds to 1 eV, the adopted criterion is actually PI$>$10\,000.} and discarded high background observing periods on the basis of an empirically derived threshold. The so-defined GTIs (Good Time Intervals) were used to produce  adequate X-ray event lists from which we extracted images and spectra. \\

For the purpose of spectral analysis, we adopted a circular extraction region centered on \hda. However, the \ngc\ cluster around this object is relatively crowded (see Fig. \ref{fig: xmmfov}) and we could not adopt an extraction radius larger than  about 40\arcsec, which corresponds to a fractional encircled energy of about 85 per cent. This extraction region was used for the three \epic\ instruments throughout  the data treatment of the observing campaign. For the same reason, we could not adopt an annular region around the source to estimate the background level. After several tests, we selected a circular region located outside the crowded part of the cluster (see Fig. \ref{fig: xmmfov}). As a complementary test, we also determined the background level using the {\sc blanksky} files created from deep field observations. These are described in the SOC technical note CAL-TN-0016-1-1\,v2.0. However these observations were performed with the {\sc thin} filter whereas our data were obtained with the {\sc thick} one. As the background sky level may be dependent on this factor, we tried to account for it: we extracted the {\sc thin} background spectra from the {\sc blanksky} files using a region identical to the extraction area adopted for the \hda\ binary. We then adjusted an empirical model to the {\sc thin} background and generated an equivalent {\sc thick} background using the {\it fake} \xspec\ command. For this operation, we used the {\it arf}+{\it rmf} response file provided by the SOC. This latter approach was limited to the \epicmos\ data sets as a proper corresponding \pn\ {\sc blanksky} file was unavailable. However,  the two background techniques give consistent results,  well within the mutual error bars. This supports the {\it a priori} idea that, due to the brightness of \hda,  the background determination is not a critical issue for this object. In the following,  we adopt the background as determined from our observations. We generated adequate {\it rmf} and {\it arf} files using the {\it rmfgen} and {\it arfgen} commands. We also used the matrices provided by the SOC. Again no difference was found between the spectra obtained in both ways. Finally, background corrected spectra were produced using the {\it grppha} command of the {\sc ftools} package.
\begin{figure}
\centering
\caption{ \label{fig: glob_lc} \epicmos\ and \pn\ light curves (in \cnts) of the \hda\ system in the 0.5-10.0\,keV energy band. Horizontal bars represent the effective duration of the exposure while vertical ones are the 1-$\sigma$ uncertainties on the count rates. }
\end{figure}
\begin{figure*}
\centering
\caption{ \label{fig: eb_lc} \epicmos\ and \pn\ light curves (in \cnts) of the \hda\ system in different energy bands. {\bf Left panel:} 0.2-0.5\,keV, {\bf Middle panel:} 0.5-1.0\,keV (\emph{upper curves}) and 1.0-2.5\,keV (\emph{lower curves}). {\bf Right panel:}  2.5-10.0\,keV. Horizontal bars represent the effective duration of the exposure while vertical ones are the 1-$\sigma$ uncertainties on the count rates. }
\end{figure*}

\section[]{\xmmepic\ Observations} \label{sect: epic_sciences}
\subsection{Light curves} \label{ssect: lc}
As a major aim of this project is to monitor the X-ray variability of the \hda\ system, we first extracted broad band light curves from the \epic\ data. Plots of the count rates against the phase (see Fig. \ref{fig: glob_lc}) clearly reveal an important enhancement of the global emission by about 60 per cent at the apastron passage. The maximum is reached at $\phi=0.66$, slightly after apastron ($\phi=0.51$), and is followed by a rapid decrease that slows down after periastron passage. 
 To refine our analysis, we selected  four energy bands: very soft (VS) [0.2-0.5\,keV], soft (S) [0.5-1.0\,keV], medium (M) [1.0-2.5\,keV] and hard (H) [2.5-10.0\,keV]. Figure \ref{fig: eb_lc} presents the background corrected light curves and reveals that the increase around apastron occurs throughout the different energy ranges. However, we outline that i) the shape of the curves in the H energy band is more symmetric around apastron, with an increase that is already present at $\phi=0.31$ and with the maximum at $\phi=0.54$; ii) the relative increase is larger for the harder energy bands. The observed enhancement indeed corresponds, respectively in the S, M and H bands, to about 50, 70 and over 100 per cent of the minimal flux level. Such an enhancement of the X-ray emission around apastron passage is definitely not compatible with a $1/D_{\mathrm{sep}}$ dependence of the flux, as is observed in other colliding wind binaries (e.g. \hdb, \citealp{Rauw02}). The corresponding hardness ratios show that the emitted X-ray flux is harder around apastron than around periastron. Finally, we emphasize that the S band accounts for about 50 to 55 per cent of the total detected counts in the [0.5-10.0\,keV] band. Spectra of \hda\ are therefore presumably rather soft.\\

\begin{figure*}
\begin{center}
% \rotatebox{-90}{
% \includegraphics[height=8cm]{fig5a.ps}}
% \rotatebox{-90}{
% \includegraphics[height=8cm]{fig5b.ps}}
% \rotatebox{-90}{
% \includegraphics[height=8cm]{fig5c.ps}}
% \rotatebox{-90}{
% \includegraphics[height=8cm]{fig5d.ps}}
% \rotatebox{-90}{
% \includegraphics[height=8cm]{fig5e.ps}}
% \rotatebox{-90}{
% \includegraphics[height=8cm]{fig5f.ps}}
\caption{ \epicmos 2 spectra of \hda\ for the different pointings of the \xmm\ satellite. 
  }  \label{fig: EPIC-MOS2 spectra}
\end{center}
\end{figure*}

\begin{table*}
\caption{
Spectral parameters of \hda\ as obtained from three-component models simultaneously fitted to the three \epic\ spectra. The upper part of this table gives the best-fit 3T models ({\tt wabs$_\mathrm{ISM}$*(wabs$_1$*mekal$_1$+wabs$_2$*mekal$_2$+wabs$_3$*mekal$_3$)}). The lower part reports the best-fit 2T+PL models ({\tt wabs$_\mathrm{ISM}$*(wabs$_1$*mekal$_1$+wabs$_2$*mekal$_2$+wabs$_3$*power)}). $n_\mathrm{H}$ gives the equivalent hydrogen column of absorbing matter (in $10^{22}$\,cm$^{-2}$), $kT$ is the temperature of the \mekal\ component (in keV) and $\Gamma$ is the spectral index of the power-law component. The normalization coefficient of the \mekal\ component  $N$ (in cm$^{-5}$) equals $\frac{10^{-14}}{4\pi d^2}\int n_\mathrm{e} n_\mathrm{H} dV$ where $d$ is the distance to the source (in cm), and $n_\mathrm{e}$ and $n_\mathrm{H}$ are the electron and hydrogen number densities (in cm$^{-3}$), whereas $N$ equals the photon flux at 1\,keV for the PL models. The upper and lower numbers quote the 90 per cent confidence intervals.
\label{tab: 3comp param}
 }
\begin{center}
\begin{tabular}{c c c c c c c c c c c c}
\hline
\multicolumn{12}{c}{Three-temperature models} \\
\hline
$\phi$ & $n_\mathrm{H,1}$ & $kT_1$  & $N_1$ (10$^{-4}$) 
       & $n_\mathrm{H,2}$ & $kT_2$ & $N_2$  (10$^{-4}$) 
       & $n_\mathrm{H,3}$ & $kT_3$ & $N_3$  (10$^{-4}$) & \chin & d.o.f\\
\hline
\vspace*{1mm}
 0.536 & $<0.03$              & $0.31^{0.32}_{0.30}$ & $9.93^{11.8}_{9.62}$ 
       & $0.48^{0.53}_{0.45}$ & $0.71^{0.72}_{0.70}$ & $11.6^{12.5}_{10.9}$
       & $<1.09$              & $4.50^{8.28}_{2.63}$ & $0.78^{1.26}_{0.56}$ & 1.68 & 419    \\
\vspace*{1mm}
 0.659 & $0.31^{0.47}_{0.13}$ & $0.18^{0.21}_{0.14}$ & $74.5^{19.9}_{205.}$ 
       & $<0.02$              & $0.53^{0.57}_{0.48}$ & $5.30^{6.50}_{4.18}$
       & $0.81^{0.95}_{0.67}$ & $0.77^{0.84}_{0.72}$ & $10.7^{12.3}_{9.19}$ & 1.49 & 314   \\
\vspace*{1mm}
 0.808 & $0.42^{1.62}_{0.38}$ & $0.14^{0.15}_{0.14}$ & $284.^{307.}_{223.}$ 
       & $<0.02$              & $0.41^{0.43}_{0.39}$ & $5.57^{5.92}_{5.23}$
       & $0.85^{0.89}_{0.75}$ & $0.80^{0.83}_{0.78}$ & $10.6^{11.4}_{10.2}$ & 1.39 & 386   \\
\vspace*{1mm} 
 0.002 & $0.29^{0.42}_{0.16}$ & $0.15^{0.17}_{0.14}$ & $105.^{285.}_{33.5}$ 
       & $<0.04$              & $0.50^{0.53}_{0.46}$ & $4.48^{8.67}_{3.96}$
       & $0.87^{1.07}_{0.72}$ & $0.89^{0.98}_{0.79}$ & $5.40^{6.94}_{4.53}$ & 1.34 & 223    \\
\vspace*{1mm}
 0.148 & $0.29^{0.39}_{0.18}$ & $0.15^{0.17}_{0.14}$ & $104.^{241.}_{53.0}$ 
       & $<0.03$              & $0.47^{0.50}_{0.41}$ & $4.05^{4.57}_{3.69}$
       & $0.78^{0.92}_{0.66}$ & $0.78^{0.86}_{0.72}$ & $5.86^{6.93}_{4.97}$ & 1.43 & 318     \\ 
\vspace*{1mm}
 0.312 & $0.38^{0.47}_{0.31}$ & $0.15^{0.16}_{0.14}$ & $182.^{375.}_{99.2}$ 
       & $<0.02$              & $0.48^{0.50}_{0.43}$ & $3.59^{3.99}_{3.19}$
       & $1.08^{1.25}_{0.97}$ & $0.83^{0.90}_{0.78}$ & $8.08^{9.45}_{6.95}$ & 1.46 & 335  \\       
\hline 
\hline
\multicolumn{12}{c}{Two-temperature + powerlaw models} \\
\hline
$\phi$ & $n_\mathrm{H,1}$            & $kT_1$               & $N_1$ (10$^{-4}$)                           
       & $n_\mathrm{H,2}$            & $kT_2$               & $N_2$ (10$^{-4}$) 
       & $n_\mathrm{H,3}$            & $\Gamma$             & $N_3$ (10$^{-4}$) & \chin & d.o.f \\
\hline
\vspace*{1mm}
 0.536 & $<0.02$              & $0.33^{0.34}_{0.32}$ & $7.73^{9.20}_{7.01}$ 
       & $0.33^{0.40}_{0.26}$ & $0.71^{0.72}_{0.69}$ & $6.59^{7.93}_{5.43}$
       & $<0.03$              & $ 3.3^{3.5}_{3.1} $  & $2.38^{2.87}_{1.81}$ & 1.56 & 419  \\
\vspace*{1mm}
 0.659 & $<0.15$              & $0.22^{0.27}_{0.20}$ & $5.54^{17.3}_{3.65}$ 
       & $<0.08$              & $0.59^{0.61}_{0.58}$ & $6.06^{7.29}_{5.17}$ 
       & $0.25^{0.63}_{0.08}$ & $ 4.1^{4.6}_{3.8} $  & $6.12^{10.1}_{4.18}$ & 1.38 & 314  \\
\vspace*{1mm}
 0.808 & $<0.06$              & $0.32^{0.33}_{0.30}$ & $7.80^{11.5}_{7.09}$ 
       & $0.43^{0.52}_{0.35}$ & $0.71^{0.73}_{0.67}$ & $5.88^{7.12}_{4.78}$   
       & $<0.03$              & $ 3.7^{3.9}_{3.5} $  & $2.57^{3.03}_{2.10}$ & 1.26 & 386  \\
\vspace*{1mm}
 0.002 & $<0.07$              & $0.20^{0.24}_{0.17}$ & $5.26^{12.1}_{4.18}$
       & $<0.04$              & $0.57^{0.59}_{0.53}$ & $4.04^{4.73}_{3.55}$
       & $<0.15$              & $ 3.6^{3.8}_{3.3} $  & $2.28^{3.01}_{1.79}$ & 1.27 & 223  \\
\vspace*{1mm}
 0.148 & $<0.33$              & $0.21^{0.26}_{0.14}$ & $5.49^{61.4}_{3.78}$
       & $<0.10$              & $0.57^{0.59}_{0.50}$ & $4.01^{5.14}_{3.19}$
       & $<0.09$              & $ 3.6^{3.8}_{3.3} $  & $2.00^{2.58}_{1.54}$ & 1.34 & 318  \\
\vspace*{1mm}
 0.312 & $<5.33$              & $0.20^{0.23}_{0.19}$ & $5.35^{34.4}_{4.56}$
       & $<0.61$              & $0.57^{0.58}_{0.54}$ & $3.07^{5.04}_{2.80}$
       & $0.10^{0.35}_{0.00}$ & $ 3.7^{3.7}_{3.4} $  & $2.93^{3.78}_{2.57}$ & 1.32 & 335  \\
\hline
\end{tabular}
\end{center}
\end{table*}

\subsection{Spectra and spectral fits}  \label{ssect: fits}
As described in Section \ref{ssect: epic_data_red}, the X-ray spectra of \hda\ were consistently extracted for the three instruments from each of the six data sets and they were binned to reach at least 25 counts per bin\footnote{We also performed the complete analysis using spectra binned with at least 10 counts per bin. This does not alter neither the qualitative conclusion, nor the quantitative values of the resulting spectral parameters of the fitted models. However, the obtained \chin\ were systematically lower than the ones obtained with 25 counts per bin spectra.}. As shown in Fig. \ref{fig: EPIC-MOS2 spectra},  the obtained spectra are relatively soft with their maximum located between 0.8 and 0.9\,keV. At first sight they reveal little variability but a general increase of the emission level after apastron passage.  In order to characterize the X-ray spectral properties of \hda\, and search for any modification with time, we investigated several spectral models using the \xspec\ v.11.0 software. In the following, we limit our study to energies above 0.5\,keV. We fixed the interstellar  column of absorbing matter to a value of $n_\mathrm{H,ISM}=0.311 \times 10^{22}$\,cm$^{-2}$ (obtained from \citet{Ryter96} formula with $A_\mathrm{V}=1.4$ \citep*{BVF99}). The fitted models are a combination of thermal \mekal\ models and powerlaw spectra and we allow a specific circumstellar column of absorbing matter for each component. For each observation, the models were adjusted to the individual \epicmos\ and \pn\ spectra. We performed simultaneous fits to all three \epic\ spectra as well. Resulting parameters are reasonably consistent and Table \ref{tab: 3comp param} gives the sole results of the \epic\ simultaneous fitting.\\

\begin{table}
\caption{
Observed (Obs.) and dereddened (Dered.) X-ray fluxes of \hda\ in the energy band 0.5-10.0\,keV corresponding to the best fitted 2T+PL models (see Table \ref{tab: 3comp param}). The dereddened fluxes were corrected for the ISM absorption only. The corresponding luminosities were computed assuming a distance modulus $DM=11.2$.
\label{tab: fluxes}
 }
\begin{center}
\begin{tabular}{l c c c}
\hline
       & \multicolumn{2}{c}{Flux (\ergs\,cm$^{-2}$)} & Luminosity (\ergs) \\
$\phi$ & Obs.             & Dered.           & Dered.           \\
\hline
 0.536 & $1.16\,10^{-12}$ & $3.11\,10^{-12}$ & $1.12\,10^{33}$ \\
 0.659 & $1.18\,10^{-12}$ & $3.33\,10^{-12}$ & $1.20\,10^{33}$ \\
 0.808 & $1.02\,10^{-12}$ & $2.96\,10^{-12}$ & $1.07\,10^{33}$ \\
 0.002 & $8.28\,10^{-13}$ & $2.58\,10^{-12}$ & $9.32\,10^{32}$ \\
 0.148 & $7.52\,10^{-13}$ & $2.36\,10^{-12}$ & $8.53\,10^{32}$ \\
 0.312 & $7.33\,10^{-13}$ & $2.23\,10^{-12}$ & $8.06\,10^{32}$ \\
\hline
\end{tabular}
\end{center}
\end{table}

It became rapidly obvious that a single temperature (1T) model was insufficient to reproduce the observed spectra ($\chi^2_{\nu}>4$). As a second step, we adopted two-component models (Fig. \ref{fig: 2comp spectra}). We either combined a 1T \mekal\ model with a powerlaw (PL) or we adjusted two-temperature (2T) \mekal\ models. Though 1T+PL models clearly give a better agreement with the data (lower \chin), they fail to reproduce the emission lines in \epic\ spectra (see e.g. the Si\,{\sc xiii} multiplet at $\sim1.85$\,keV in Fig. \ref{fig:  2comp spectra}). Restraining the fit to the spectral region below 2.5\,keV considerably improves the quality of the 2T models with respect to the 1T+PL models. Actually the power law component helps to reproduce the flux at high energy that is not explained by 2T models in some of the spectra that show enough signal above 3\,keV. Adding a PL as a third component to the 2T models again increased the quality of the fit for those particular spectra (Fig. \ref{fig: 3comp spectra}). The high energy tail could also be reproduced by a high temperature ($kT_3 \sim 4$\,keV)  \mekal\ component. However, except for Obs. 1 ($\phi=0.54$), a slightly better $\chi^2$ is reached using 3T models for which the three components are soft ($kT<1$\,keV). In this latter case however, the hard energy tail might sometimes not be adequately reproduced. Though 2T+PL models provide slightly better fits, it is however tricky to definitely choose among all these types of models. If the high energy tail does actually correspond to a PL component, this latter is then characterized by a photon index $\Gamma$\,$\simeq$\,3.7. This component has a much steeper slope than the value $\Gamma$\,$=$\,1.5 that is expected from inverse Compton scattering emission produced by a relativistic population of electrons accelerated in strong shocks \citep{CW91}. Table \ref{tab: fluxes} provides the observed and dereddened fluxes as deduced from the best-fit 2T+PL models.\\

\begin{figure*}
\begin{center}
% \rotatebox{-90}{
% \includegraphics[height=8cm]{fig6a.ps}}
% \rotatebox{-90}{
% \includegraphics[height=8cm]{fig6b.ps}}
% \rotatebox{-90}{
% \includegraphics[height=8cm]{fig6c.ps}}
% \rotatebox{-90}{
% \includegraphics[height=8cm]{fig6d.ps}}
\caption{\epicmos1 and \pn\ spectra at $\phi=0.54$ (upper panels) and $\phi=0.15$ (lower panels). The spectra were fitted either with 2T \mekal\ models (lefthand column) or 1T+PL models (righthand column). The model components are drawn with dashed or dashed-dotted lines while the solid lines represent the resulting models. Note how the PL component compensates the flux above 3\,keV, resulting in a fitted model that does not provide a satisfying fit to the emission lines.
\label{fig: 2comp spectra}}
\end{center}

\end{figure*}

\begin{figure*}
\begin{center}
% \rotatebox{-90}{
% \includegraphics[height=8cm]{fig7a.ps}}
% \rotatebox{-90}{
% \includegraphics[height=8cm]{fig7b.ps}}
% \rotatebox{-90}{
% \includegraphics[height=8cm]{fig7c.ps}}
% \rotatebox{-90}{
% \includegraphics[height=8cm]{fig7d.ps}}
\caption{\epicmos 1 and \pn\ spectra at $\phi=0.54$ (upper panels) and $\phi=0.31$ (lower panels). The spectra were fitted either with 3T \mekal\ models (lefthand column) or 2T+PL models (righthand column). The model components are drawn with dashed, dashed-dotted or dotted lines while the solid lines represent the resulting models. Note the improvement compared to two-component models.
\label{fig: 3comp spectra}}
\end{center}
\end{figure*}

\begin{table*}
\caption{
Spectral parameters of \hda\ as obtained from 2T \mekal\ models simultaneously fitted to the three \epic\ spectra in the range 0.5-2.5\,keV. Details of the models are: {\tt wabs$_\mathrm{ISM}$*(wabs$_1$*mekal$_1$+wabs$_2$*mekal$_2$)} in which the value of $n_\mathrm{H,1}$ has been fixed to zero (see text). 
The same notations as in Table \ref{tab: 3comp param} have been used.
\label{tab: 2t}
 }
\begin{center}
\begin{tabular}{c c c c c c c c c}
\hline
$\phi$ & $n_\mathrm{H,1}$ & $kT_1$ & $N_1$   
       & $n_\mathrm{H,2}$ & $kT_2$ & $N_2$   & \chin & d.o.f\\
       &(cm$^{-2}$)          & (keV)  & (10$^{-4}$\,cm$^{-5}$)
       &(10$^{22}$\,cm$^{-2}$) & (keV)  & (10$^{-4}$\,cm$^{-5}$) \\
\hline
\vspace*{1mm}
0.536 &  0                    &  $0.310^{0.317}_{0.303}$ & $10.12^{10.44}_{9.797}$ 
      &  $0.54^{0.58}_{0.51}$ &  $0.712^{0.726}_{0.699}$ & $13.61^{14.24}_{12.97}$  & 1.73 & 390 \\
\vspace*{1mm}
0.659 &  0                    &  $0.304^{0.316}_{0.289}$ & $10.62^{11.11}_{10.08}$ 
      &  $0.48^{0.53}_{0.43}$ &  $0.694^{0.716}_{0.657}$ & $13.57^{15.05}_{12.43}$  & 1.51 & 302 \\
\vspace*{1mm}
0.808 &  0                    &  $0.293^{0.302}_{0.284}$ & $10.51^{10.87}_{10.17}$ 
      &  $0.59^{0.63}_{0.45}$ &  $0.694^{0.731}_{0.663}$ & $12.56^{13.71}_{11.40}$  & 1.46 & 363 \\
\vspace*{1mm}
0.002 &  0                    &  $0.257^{0.271}_{0.239}$ & $9.057^{9.639}_{8.431}$ 
      &  $0.41^{0.47}_{0.34}$ &  $0.623^{0.648}_{0.595}$ & $9.303^{10.30}_{8.385}$  & 1.47 & 216 \\
\vspace*{1mm}
0.148 &  0                    &  $0.256^{0.268}_{0.243}$ & $8.734^{9.151}_{8.287}$ 
      &  $0.42^{0.48}_{0.36}$ &  $0.621^{0.648}_{0.599}$ & $8.769^{9.511}_{8.085}$  & 1.45 & 308\\
\vspace*{1mm}
0.312 &  0                    &  $0.272^{0.290}_{0.258}$ & $8.127^{8.584}_{7.748}$ 
      &  $0.60^{0.65}_{0.53}$ &  $0.656^{0.709}_{0.625}$ & $9.183^{10.71}_{8.620}$  & 1.53 & 317\\
\hline
\end{tabular}
\end{center}
\end{table*}

Though the X-ray flux from \hda\ definitely shows a clear variation by about a factor 1.6 between its lower and higher value, the  spectra reveal very little variability at first sight. To detect and to quantify the evolution of the fitted spectral parameters in the different pointings is not straightforward as different models, but also different solutions of the same model, fit the data with a similar quality. This latter aspect can be seen from Table \ref{tab: 3comp param} in which the values of the spectral parameters may vary discontinuously from one phase to another. This indeterminacy probably results from too large a number of free parameters and from the lack of physical constraints imposed on the model parameters. We therefore choose to limit the fitted interval to the range [0.5 - 2.5\,keV] which is reasonably well reproduced using a 2T \mekal\ model. This approach offers the advantage of avoiding the uncertainty about the nature of the hard-energy tail. Though this helps to improve the situation, some of the fits still present local minima of similar depth. Best fitted values for the column of absorbing matter of the first component ($n_\mathrm{H,1}$) usually tend to be very small, if not equal to zero. Adopting 2T models where this parameter is neglected ($n_\mathrm{H,1}=0$\,cm$^{-2}$) improves slightly but systematically the quality of the different fits. It also solves the degeneracy between the local minima of the fit, resulting in a coherent fit for the six \xmm\ pointings. Results of these fits are presented in Table \ref{tab: 2t}. The fact that the column of absorbing matter of the cold component of the model is ill-constrained  could be due to the fact that the fitted parameter $n_\mathrm{H}$ stands for an equivalent hydrogen column density of \emph{neutral}\ gas. The material that is responsible for the local absorption of the X-ray flux in the \hda\ system, however, is ionized due to the amount of UV photons emitted by the two O-type stars. A neutral absorbing column can then probably not properly reproduce the local circumstellar absorption. The resulting discrepancies will mainly manifest themselves at lower energies where the absorption is the most efficient.\\

Figure \ref{fig: 2t} shows the evolution, with the phase, of the best-fit values of the 2T \mekal\ parameters in the range 0.5-2.5\,keV. The temperatures and normalization coefficients of the two components present clear variations. These are in fair agreement with the observed modulations of the broad band light curves (see Fig. \ref{fig: eb_lc}). Variations of the equivalent column of absorbing matter of the higher temperature component are less clear cut. However, there might exist two increases at phases $\phi=0.3$ and $\phi=0.8$ which roughly correspond to the quadrature phases of the \hda\ system. Though the confidence intervals are quite large, this could indicate an enhancement of the local absorption while the line of sight is almost perpendicular to the binary axis and therefore crosses the radial structure of the interaction region (see Sect. \ref{sect: simul}). \\

\begin{figure*}
\centering
\caption{ \label{fig: 2t} Best-fit spectral parameters of the 2T \mekal\ models ({\tt wabs$_\mathrm{ISM}$*(wabs$_1$*mekal$_1$+wabs$_2$*mekal$_2$)}) in the range 0.5-2.5\,keV plotted vs. the phase. Open diamonds refer to the first, low temperature, component parameters while filled diamonds stand for the second temperature component (see Table \ref{tab: 2t}). {\bf Left:} Equivalent column of neutral hydrogen $n_\mathrm{H}$. {\bf Centre:} Temperature $kT$. {\bf Right:} Normalization coefficient $N$. 
}
\end{figure*}

\begin{figure}
\centering
\caption{Comparison of observed \rosatpspc\ count rates (open symbols) and equivalent \xmmnew\ count rates (filled symbols) of \hda\ in the range [0.2-2.4\,keV]. Phases of the \rosat\ observations have been computed adopting the ephemerides reported in Table \ref{tab: orbit}. 
\label{fig: rosat_flux}
}
\end{figure}

\subsection{Phase-locked variability} \label{ssect: phase-locked}
Though the observations clearly show a decrease of the emitted flux from Obs. 1-2 to Obs. 6, our \xmm\ campaign alone can however neither establish the phase-locked behaviour of the observed variations, nor whether this behaviour is stable on longer time-scales. Indeed, as seen from Table \ref{tab: journal}, the \xmm\ observations of \hda\ extended over about 5 days and therefore did not cover more than a single orbital cycle of the binary. In order to investigate this question, we retrieved previous \rosatpspc\ observations of the \ngc\ cluster \citep{Corc96} and we re-analysed the data adopting the ephemerides from Table \ref{tab: orbit}, that are much better constrained than the ones \citeauthor{Corc96} used at the time. We extracted light curves and spectra in the ROSAT energy band. Due to the modest sensitivity and energy coverage of the satellite, the \pspc\ spectra are well reproduced by a single temperature component at $kT=0.24$\,keV and no significant spectral variability could be observed in the \pspc\ data. 
With the help of the \xspec\ software, we then combined the 2T+PL models that best reproduced the \xmm\ observations with the \pspc\ response file to estimate the equivalent \pspc\ count rates of our \xmm\ data. Figure \ref{fig: rosat_flux} gives a direct comparison of the equivalent \pspc\ rates with the actual \rosat\ observations in the same energy range. In view of the difficulty to accurately compare the count rates measured from two such different satellites, the current agreement between \xmm\ and \rosat\ observations is quite reasonable and gives a strong support to a phase-locked behaviour of the observed variability. It also suggests the relative stability, on longer time scales, of the dominant phenomenon that produces the observed variation in the light curves of \hda. Indeed, the \rosatpspc\ observations were obtained between March 1991 and February 1993, i.e. about 10 years prior to our \xmm\ monitoring campaign. \\

\section{Hydrodynamical simulations} \label{sect: simul}
As suggested by \citet{PS97}, comparisons of X-ray observations of CWBs with detailed hydrodynamic simulations may help to better understand and constrain the physics of the phenomenon. Indeed our \xmmnew\ campaign reveals precious information about the global flux emitted by the system that {\it a priori} results from the X-rays produced in the interaction area as well as within the denser layers of the winds of the two O-type components. However without further analysis they provide little understanding of the local physics of the interaction region. Indeed, the `inversion' of the X-ray data to recover the details of the local emission processes as well as the various parameters that affect them is certainly a very ill-posed problem. In this regard, Computational Fluid Dynamics (CFD) has proved to open a new window in the physics of the X-ray emission from the winds themselves \citep{Feld97,DO03} and from the wind-wind collision zone \citep{PS97,PS02}. \\

This section is organized as follows. First, we give a general description of the code used to solve the hydrodynamic problem and we describe the performed computational runs (Sect. \ref{ssect: hcode}). Sect. \ref{ssect: ecode} presents the model used to estimate the fluxes that are predicted by the hydrodynamical grids. We then report the predictions of the present model (Sect. \ref{ssect: predict}) and we compare them to the \xmm\ observations (Sect. \ref{ssect: th_obs_compar}). Finally, we conclude by discussing some of the obvious limitations of this work (Sect. \ref{ssect: limit}).\\

\subsection{The hydrodynamical code} \label{ssect: hcode}
In this section, we limit ourselves to a brief overview of the hydrodynamical code. A more complete description is given by SBP92 and references therein. The VH-1 hydrodynamic code aims at the resolution of the Euler differential set of equations for inviscid fluids. It is based on the Piece-wise Parabolic Method \citep{CW84}, a third-order accurate time-marching algorithm that is proved to be shock-capturing and therefore naturally takes care of the shocks. Very briefly, this method is a generalisation of the Gudonov scheme that handles each cell interface as a shock-tube problem. In this regard, it builds the global solution of the equations by solving the local behaviour of the fluid \citep[see][ for a general introduction to CFD techniques]{And95}. \\

\begin{table}
\centering
\caption{Effective separation and wind velocities at the ram pressure equilibrium  surface on the axis of the system (see text) as a function of the \hda\ orbital phase  $\phi$ at the time of the six \xmm\ observations. The last column provides the logarithm of the  total theoretical X-ray luminosity computed for an idealized planar collision region (see text).
\label{tab: windvel}}
\begin{tabular}{c c c c c c c}
\hline
Obs. & $\phi$ & \multicolumn{2}{c}{Separation} & \multicolumn{2}{c}{Wind vel. (\kms)} & $\log L^\mathrm{th}_\mathrm{X}$ \vspace*{1mm}\\
\# &     & \rsol & $10^{12}$\,cm & O7\,III & O7.5\,III & \ergs \\
\hline
1 & 0.54 & 59.8  & 4.16  & 1240  &  1330 & 35.73 \\ 
2 & 0.66 & 57.4  & 3.98  & 1180  &  1280 & 35.70 \\ 
3 & 0.81 & 51.5  & 3.58  &  990  &  1160 & 35.58 \\ 
4 & 0.00 & 46.4  & 3.22  &  950  &  1150 & 35.56 \\ 
5 & 0.15 & 49.9  & 3.46  &  910  &  1160 & 35.55 \\
6 & 0.31 & 56.4  & 3.92  & 1160  &  1260 & 35.68 \\
\hline
\end{tabular}
\end{table}

In this work, we adopt the simplified configuration of two spherically symmetric winds of constant velocities. We therefore both  neglect the orbital motion and the wind acceleration. We assume that the wind material is fully ionized ($\gamma=\frac{5}{3}$) and behaves as a perfect gas. Our computations account neither for the viscosity, nor for a possible magnetic field. These approximations  result in an axi-symmetrical geometry around the line of centres and allow to reduce the hydrodynamical problem to a two-dimensional flow. For that reason, all the hydrodynamical runs presented in this paper are performed on 2-D grids. We adopt a typical grid size of $300 \times 600$ cells, corresponding to a physical size of $6\,10^{12} \times 10^{13}$\,cm, and we let the flow evolve with time from an initial situation where both winds have not yet collided. We let the interaction grow while the time elapses and we follow the process for at least 4000 steps, long enough so that a \emph{steady state} may be reached. However, the \hda\ CWB harbours a highly radiative collision zone and the post-shock flow is dominated by instabilities. Therefore no `true' steady state may be reached. Nevertheless the chosen evolution time of the simulations is long enough for the flow to relax from the initial condition dependencies. The time elapsed between two iterations is a free parameter that is adjusted at each step according to numerical stability considerations, but a typical time interval between two steps is about 70 to 120 sec. We performed test runs with higher- as well as lower-resolution grids to check that the observed behaviour of the flow is not, within the tested configurations, resolution-dependent (see however Sect. \ref{ssect: limit}).
Energy-loss due to radiation is computed after each hydrodynamical iteration and is the only source term accounted for in the adopted version of the code.  
Reflective boundary conditions were imposed along the line of star centres while we let the gas flow freely out of the grid at the other three boundaries.\\

As our aim is to compare observations and model predictions, the hydrodynamical runs were performed adopting a system configuration and wind parameters that  closely reproduced the configuration of the \hda\ system at the time of the six \xmm\ observations. Table \ref{tab: windvel} reports the adopted values for the physical parameters. The separations between the stars were computed using the orbital parameters and ephemerides tabulated in Table \ref{tab: orbit}. Mass losses are deduced from \citet*{VdKL00} formula with terminal wind velocities $v_{\infty}=2420$\,\kms\ \citep{HSHP97} as quoted in Paper I\footnote{In Paper I, we reported a slightly erroneous value for the terminal velocity of the secondary component. The present value (see Table \ref{tab: orbit}) is therefore corrected accordingly. Fortunately, this does not alter the mass loss rate computation. Our previous conclusions remains therefore unchanged except that, using the present corrected values, the on axis wind momentum ratio is now slightly in favour of the secondary star.}. \hda\ being a close binary, the winds collide well before reaching their terminal velocities. As wind acceleration is neglected, we circumvent this problem and adopt the wind velocities reached at the position of the ram pressure equilibrium surface on the binary axis. This is obtained by solving the on-axis ram pressure equilibrium equation: 
\begin{equation} 
\frac{\dot{M}_1 v_1}{d^2_1}=\frac{\dot{M}_2 v_2}{d^2_2} \label{eq: ram pressure}
\end{equation}
where $d_1$ and $d_2$ are the distances from the equilibrium surface to either star centre; $d_1+d_2$ is the separation between the centres. Both wind  velocities $v_1$ and $v_2$ are computed from a standard $\beta$-law with $\beta=0.8$. Equation \ref{eq: ram pressure} usually provides three solutions, two of which correspond to unstable equilibria. The third one gives the position of the stable equilibrium that we then use to derive the constant wind velocities reported in Table \ref{tab: windvel}. Though this  is a rather crude method, it allows to merely bypass the problem of non-terminal wind velocities. This provide a reasonable approximation to the fact that, as the system is rather eccentric,  the orbital separation is variable and the winds therefore reach quite different velocities at the shock surfaces for the different orbital phases that we are studying.\\

\begin{figure*}
\begin{center} 
\caption{Density (left) and temperature (right) maps as computed from our hydrodynamical simulations. The adopted parameters (see text) reproduce the \hda\ configuration at phase $\phi = 0.15$.
\label{fig: hrsp5}}
\end{center}
\end{figure*}

The hydrodynamical runs performed adopting these six sets of parameters finally provide the evolution of the values of the gas density, gas pressure and the radial and axial velocity components at each grid cell, from which other hydrodynamical variables (such as temperature, internal energy, entropy...) and radiative properties (see Sect. \ref{ssect: ecode}) can be inferred. Figure \ref{fig: hrsp5}  gives typical density and temperature maps provided by the simulations of the \hda\ CWB system. \\

\subsection{X-ray emission} \label{ssect: ecode}
From the results of the simulations described in the previous section, we extracted maps of the hydrodynamical configuration (i.e. density, pressure, temperature, r- and z-velocity component maps ...) once every 10 time steps. Depending on the initial configuration, it takes about 2100 to 2600 iterations for the system to relax from the initial conditions, which corresponds to an elapsed time ranging from about 1.5 to 2.0 10$^5$\,sec. Once the dynamical flow has relaxed, we obtained a snapshot approximately every 700 to 750\,sec. From the succession of these snapshots, we thus follow the evolution of the hydrodynamical variables with time, and therefore of the wind-wind interaction phenomenon, in the {\it frozen} configuration of the \hda\ system that corresponds to one of the six \xmm\ observations. By themselves, the hydrodynamical simulations are very instructive and it is worth to monitor the flow structure of the collision area. However, to compare these simulations with observational results, we need to deduce observable variables from the hydrodynamical grids. For this purpose we then solve the radiative problem corresponding to the different hydrodynamical configurations.\\

The X-ray emission from the system was evaluated by summing up the emissivity of each cell, accounting for a proper extinction (see details in SBP92). The computation of the column of absorbing matter along the line of sight required a three-dimensional geometry as we have  to account both for the inclination of the system and for the variable line of sight of the observer. A 3-D grid was thus generated by a rotation of the 2-D hydrodynamical grid around the line of centres. The adopted angular resolution of the constructed 3-D cells is 2\degr. The chemical composition of both winds was supposed to be solar. For each cell, we computed the emissivity as a function of the mean cell temperature. The contribution of grid cells with a temperature lower than $10^6$\,K was neglected. For the emitting cells, the corresponding column of circumstellar absorbing matter towards the observer was computed and we used proper absorbing coefficients to derive the intrinsic attenuated spectra. Cells with temperature higher than $10^9$\,K were considered not to give a significant contribution to the absorption. Eclipses were accounted for assuming stellar radii of 15 \rsol. Interstellar absorption was taken into account assuming a column of absorbing matter of $n_\mathrm{H,ISM}=0.311 \times 10^{22}$\,cm$^{-2}$. We finally obtained predicted X-ray fluxes and spectra that are suitable for a direct comparison with the \xmm\ observations. \\

\begin{figure}
\begin{center} 
\caption{ Cross-section at $z=2\,10^{12}$\,cm in the density (dashed line) and temperature (solid line) maps presented in Fig. \ref{fig: hrsp5}. Both variables are plotted using logarithmic scales.
\label{fig: dens_temp}}
\end{center}
\end{figure}

\subsection{Predictions of the simulations} \label{ssect: predict}
As can be seen from Fig. \ref{fig: hrsp5}, the  collision zone turns out to  be highly unstable. The wind-wind interaction region appears to be extremely thin and to develop filaments that form smoke-like structures while time elapses. Similar patterns are found for the six configurations in which we performed the simulations. The thinness of the collision zone is expected due to the highly radiative behaviour of the collision ($\chi=0.02-0.15$). The interaction region  appears to be so narrow in our simulations that its inner structure could not be resolved with our mesh size. Increasing the resolution by a factor two does not solve the problem. This failure to resolve the collision area constitutes an obvious limitation of our work that will be discussed in Sect. \ref{ssect: limit}.\\

The filaments that form the interaction region slowly evolve to produce thin but dense shells that are separated by low density regions (Fig. \ref{fig: dens_temp}). In addition to this general pattern, some pockets of high temperature gas seem to undergo an adiabatic expansion with typical cooling time scale larger than a couple of weeks. It is probable that some of these pockets might survive till far from the axis but, due to their very low density, it is unlikely that they provide a significant contribution to the total emission from the wind-wind interaction.\\

To investigate the collision zone luminosity, we computed the expected emission for each of the extracted snapshots. One of the surprising results of these simulations is that the luminosity from the wind collision can  vary by a factor of a few on time scales of a couple of hours (see Fig. \ref{fig: lc_compar}). As the time resolution of our simulations is about 750\,sec, we also performed a run in a particular configuration, extracting snapshots after each hydrodynamical iteration. This provides a light curve with a time step of about 75\,sec. The latter almost exactly matches the previous light curve that has a resolution ten times lower. We can therefore reasonably assume that all the temporal information concerning the variability of the luminosity are already contained in the light curves presented in Fig. \ref{fig: lc_compar}.
To estimate the expected luminosity at a given phase of \hda, we therefore computed the average and dispersion of the luminosity of the selected snapshots.  This is presented in Fig. \ref{fig: prdcted_lum} where the vertical bars should be considered as the model envelope rather than as `classical' error bars. The upper panel gives the variation of the intrinsic emission produced in the collision of the winds. The latter displays a variation of about 40 per cent between periastron and apastron. However accounting for the eclipses and the absorption by the circum- and interstellar medium drastically changes the relative amplitude of the variation. From the lower panel of Fig. \ref{fig: prdcted_lum}, it is obvious that the performed hydrodynamical simulations predict a phase-locked variation of the outgoing flux by about one order of magnitude. The maximum of the flux is reached around apastron, while the emitted flux is minimum at periastron. \\

\begin{figure}
\begin{center} 
\caption{Evolution with the time of the logarithm of the absorbed X-ray luminosity emerging from the interaction region in the range 0.5-10.0\,keV, as predicted by the models. Different lines refer to different configurations of the system. Solid line: $\phi=0.54$. Dashed line: $\phi=0.00$. Dotted line: $\phi=0.15$.
\label{fig: lc_compar}}
\end{center}
\end{figure}

In order to check the quality of our model, we also computed the theoretical luminosity emitted by the collision of two winds of equal strength that would form an idealized planar interaction region. In this simple model, we further assumed that the kinetical energy of the flow component radial to the shock surface is fully converted into X-rays. For each wind, this is given by \citep*{LMM90}: 
\begin{equation}
L_{\mathrm{X,}i}^\mathrm{th}=\frac{L_{\mathrm{w,}i}}{6}=\frac{\dot{M}_{i} v^2_{i}}{12}
\end{equation}
where $L_{\mathrm{w,}i}=\frac{\dot{M}_{i} v^2_{i}}{2}$ is the equivalent wind luminosity of star $i$, characterized by a mass-loss rate $\dot{M}_{i}$ and a pre-shock velocity $v_{i}$. Typical values for the kinetic energy are of the order of $10^{36}-10^{37}$ \ergs\ for single O-stars with terminal wind velocities. Values reported in Table \ref{tab: windvel} for the \hda\ system result from the sum of the X-ray luminosities from the two shocked winds : $L_\mathrm{X}^\mathrm{th}=L_\mathrm{X,1}^\mathrm{th}+L_\mathrm{X,2}^\mathrm{th}$. These values are in excellent agreement with the averaged intrinsic luminosities predicted by the hydrodynamical simulations  (see Fig. \ref{fig: prdcted_lum}) and lend further support to the consistency of our numerical results.\\

\begin{figure}
\begin{center} 
\caption{Simulated X-ray luminosity from the collision zone in the range 0.5-10.0\,keV as predicted by the models at the different phases of \hda. \emph{Upper panel:} Open circles: Intrinsic emission of the collision zone. Filled triangles: theoretical X-ray luminosities from the collision of two identical winds (see Table \ref{tab: windvel}). \emph{Lower panel :} Emerging luminosity accounting for i) eclipses, ii) local absorption, iii) ISM absorption. The vertical bars represent the 1-$\sigma$ dispersion on the emerging fluxes.
\label{fig: prdcted_lum}}
\end{center}
\end{figure}

\subsection{Comparison with the \xmmnew\ observations and discussion} \label{ssect: th_obs_compar}
The hydrodynamical simulations of wind-wind collisions performed in configurations as close as possible to the ones of the \hda\ system at the different phases of interest predict an increase of the emitted fluxes from the interaction region by about one order of magnitude around apastron. Though the observed \epic\ light curves indeed reveal an enhancement of the emission at this particular phase, the increase is limited to a factor of about 1.6 and further shows a clear asymmetry between the phase interval [0.5-1.0] and [0.0-0.5]. Due to the 2-D nature of the simulations, the predicted light curve is naturally symmetric around apastron and, in this regard, does not render the observations. This might suggest that the observed asymmetry could partly find an explanation in the orbital motion and the resulting  deflection of the colliding wind zone. \\

\begin{figure}
\begin{center} 
\caption{Comparison between the computed dereddened luminosities as resulting from our simulation runs and the observational dereddened luminosities. Filled squares: dereddened luminosities of the interaction region as predicted by the model. The vertical bars have the same meaning as in Fig. \ref{fig: prdcted_lum}. Filled circles: total predicted luminosities, i.e. resulting from the predicted emission of the interaction region plus the expected contribution from the two components of \hda\ (see text). Open circles: observational dereddened luminosities as deduced from the fitting of the \xmmepic\ spectra (see Table \ref{tab: fluxes}).
\label{fig: th_obs_compar}}
\end{center}
\end{figure}

Assuming a distance modulus $DM=11.2$ \citep{RCB97} as the one we used to infer the bolometric luminosities and subsequent mass-loss rates reported in Paper I and Table \ref{tab: orbit}, the predicted  model luminosity clearly underestimates the `true' luminosity of the system (see Table \ref{tab: fluxes}). However the luminosities computed from the hydrodynamical simulations  only account for the emission coming from the wind-wind collision itself. Until now, the intrinsic X-ray emission of the two O stars has indeed been neglected. From Paper I, we know that $\log(L^\mathrm{prim}_\mathrm{bol}/L_{\odot})$ and $\log(L^\mathrm{sec}_\mathrm{bol}/L_{\odot})$ equal 5.61 and 5.63  respectively  for the primary and secondary component. Using the relation of \citet{Berg97} for stars with $L_\mathrm{bol}>10^{38}$\ergs :
\begin{equation}
\log\left(L_\mathrm{X}\right) = 1.13 \log\left(L_\mathrm{bol}\right)-11.89
\end{equation}
 we find that the intrinsic X-ray luminosities of the components of \hda\ should be  about $\log(L^\mathrm{prim}_\mathrm{X})=32.39$ and $\log(L^\mathrm{sec}_\mathrm{X})=32.42$ (\ergs) respectively. Adding the dereddened flux originating from the interaction region therefore provides a total X-ray luminosity suitable for a comparison with the observed values. We outline that the \citeauthor{Berg97} relation is based on observed X-ray emissions from O-stars and so includes the effect of intrinsic wind absorption. The resulting luminosities of the components of \hda\ are however not corrected for the additional absorption associated with the colliding wind region nor for the eclipses that occur in the system.   
Figure \ref{fig: th_obs_compar} presents a direct comparison of the dereddened observational X-ray luminosities with the luminosities predicted by the simulations. The latter are obtained by summing up the  emission from the interaction region with the expected intrinsic contributions from both stars of the system. Though there is still a difference of about a factor two between the observations and the model for some of the pointings, the agreement is now much better. The intrinsic dispersion of the \citeauthor{Berg97} relation and the resulting uncertainty on the intrinsic $L_\mathrm{X}$ can, by itself, be responsible for such a discrepancy. The amplitude of the variation between apastron and periastron in the model and in the observed light curve differs by about 50 per cent. Given the approximations inherent to our simulations, the agreement can be considered as quite good. \\

\begin{figure}
\begin{center} 
\caption{Comparison between the computed spectra as resulting from our simulation runs and the observational best-fit model for the configuration of the \hda\ system at $\phi=0.31$. Plain thin line: intrinsic predicted spectrum. Dashed line: absorbed predicted spectrum. Vertical bars stand for the model envelope due to fluctuations in the predicted spectrum. Thick line: best-fit 2T+PL model at $\phi=0.31$ (see Table \ref{tab: 3comp param}). 
\label{fig: spec_compar}}
\end{center}
\end{figure}

Though the quality of our model is limited (see Sect. \ref{ssect: limit}), these results suggest that the observed phase-locked behaviour of the X-ray flux of \hda\ may indeed be reasonably explained by a variation of the X-ray emission produced in the interaction region. While the intrinsic colliding wind flux increases at apastron due to larger pre-shock wind velocities, the major effect is to be attributed to the variation of the absorption properties. The predicted modulation, if diluted by the intrinsic emission of the two components of the system, indeed reasonably reproduced the observed amplitude of the fluxes deduced from the \xmmepic\ spectra.\\

The energy distribution of the predicted X-ray emission presents its maximum around 3\,keV. Figure \ref{fig: spec_compar} provides a direct comparison with the best-fit model 2T+PL that reproduces the observational spectral properties of the \hda\ system. Similar patterns are found in the six configurations. It is seen that the predicted spectra are much harder than the observed ones and, on average, underestimate the flux at low-energy (below 2\,keV) while the flux is overestimated at higher energies. However, as previously emphasized,  the predicted spectra only account for the wind-wind collision emission and the X-ray fluxes from the two stars are neglected. The X-ray emission from individual stars is expected to be relatively soft, with its maximum peak being reached around 1\,keV. This can therefore account for the observed discrepancy at low energy. On the other hand, the disagreement of the high-energy tail might come from an excessive strength of the shocks in the model, possibly resulting from too large pre-shock wind velocities being assumed. Wind acceleration might indeed be slowed down by specific effects such as radiative inhibition \citep{SP94}. However, the boundaries of the spectra predicted by the hydrodynamical models still overlap the observational best-fit model at high-energy. \\

Finally, a moderately strong 6.7\,keV Fe line is clearly predicted by the simulations while its presence is not so obvious in the data presented until now. The observational model plotted in Fig. \ref{fig: spec_compar} is a 2T+PL model in which the PL component accounts for the high energy tail and could intrinsically not reproduce the Fe\,{\sc xxv} line at  6.7\,keV. 
Determining whether this line is present or not in the spectrum of \hda\ might help to clarify the nature of the hard energy tail. 
 For this purpose, we combined the \epic\ data of the six \xmm\ pointings to increase the signal-to-noise ratio at high energy. We built merged event lists and created combined \mos\ and \pn\ {\it arf} using the {\sc ftools} command {\it addarf}. The merged \pn\ spectrum of \hda\ is presented in Fig. \ref{fig: fe_line} and corresponds to a total exposure time around 130\,ksec. We then adjusted an empirical model ({\tt power+gauss}) in the range [4.0-10.0\,keV]. This model is made of a PL component, that reproduces the continuum, and of a Gaussian component ($\frac{K}{\sqrt{2\pi}\sigma_{\mathrm{G}}} \exp \left( -(\frac{E-E_{\mathrm{G}}}{\sigma_{\mathrm{G}}})^2 \right)$) that fits the Fe line (see Fig. \ref{fig: fe_line}). Best-fit values for the Gaussian parameters give $E_{\mathrm{G}}=6.70 \pm 0.14$\,keV, $\sigma_{\mathrm{G}}=0.24 \pm 0.14$\,keV and $K=4.3 \pm 2.6\,10^{-7}$\,\cnts\,cm$^{-2}$, yielding an equivalent width of about 2.3\,keV.
From the merged data, we can infer that the 6.7\,keV Fe line is most probably present in the \hda\ spectrum and that, therefore, at least part of the high energy tail has a thermal origin. Finally, we mention that \citet{Deb04} recently noted that non-thermal phenomena could also contribute to the Fe\,{\sc xxv} emission at 6.7\,keV. The above diagnostic seems however to remain quite robust as these authors mentioned that the relative contribution of the non-thermal and thermal processes to the emission is directly proportional to the fraction of the free electron population that reaches relativistic energies. It is therefore to be expected that a non-thermal contribution to the Fe\,{\sc xxv} line remains marginal.\\

\begin{figure}
\begin{center} 
% \rotatebox{-90}{
% \includegraphics[height=8.3cm]{fig16.ps}}
\caption{Merged \epicpn\ spectrum of the \hda\ binary. The spectrum is plotted using 17 counts per bin for the sake of clarity. The inset presents the result of a PL + Gaussian model adjusted in the vicinity of the 6.7\,keV\,Fe line (see text). 
\label{fig: fe_line}}
\end{center}
\end{figure}

\subsection{Limitations of the model} \label{ssect: limit}
A first obvious limitation of these simulations is the use of constant wind velocities in a geometry for which we definitively know that the collision occurs in the acceleration region of the winds. Adopting the velocities reached at the ram pressure equilibrium surface offers nevertheless the advantage of a quite straightforward solution. This approach probably provides a good approximation close to the line of centres. However, our simulations reveal that the structure of the interaction extends far from the binary axis and the hypothesis of constant wind velocity is definitely too simplistic there. One of the ways to account for the line-driven acceleration would be to include an extra source term that would express a radiative force on the wind material. Due to the wind acceleration, the pre-shock wind velocities away from the axis would definitely be larger and the shocks stronger. Under the influence of an increased ram pressure, and of its component normal to the shock surface, the interaction region could be more confined along the radial direction. However such an effect is difficult to estimate and, furthermore, would probably  be disrupted by the orbital motion in the system.\\

Moreover, as seen from Fig. \ref{fig: hrsp5}, our grid is too coarse to resolve the inner structure of the interaction region. Though a thin interaction is expected due to the importance of the radiative cooling, a linear increase of the resolution would however not have helped to solve the problem. Indeed we are facing the need to simultaneously resolve small scale structures around and inside the interaction region, but also to follow the evolution of large scale structures such as the shells. However computation time prevents us from increasing both the resolution of the mesh and simultaneously maintaining a good coverage of the collision far from the axis. From the density and temperature conditions of the interaction region inferred from our simulations, we estimate the typical cooling length of the shock-heated plasma to be around $10^7-10^8$\,cm. On axis however, extreme values of a few $10^6$\,cm can be found. Our grid resolution along the axis is $2\,10^{10}$\,cm and is definitely not designed to resolve the inner structure of the collision zone. Increasing our resolution by a factor of $10^4$, is probably required to overcome that problem. This is unfortunately not realistic with the present hardware resources.
In such a situation, the need to adopt non-linear metrics that would increase the resolution towards the central axis of the systems is obvious. An adaptative grid that automatically clusters grid points in the region of high flow-field gradients \citep[see e.g.][]{And95} certainly offers another powerful way to overcome this problem. Due to the instability of the region and the complex structure that results from it, the mesh should also ideally be readjusted after each iteration. Therefore, to adequately perform hydrodynamical simulations of wind-wind collisions in close binary systems, an adaptative mesh algorithm, able to identify the regions where an increased resolution is needed, can provide an adequate description using reasonable time and resources. The implementation of such techniques is not trivial, but remains one of the most promising way to drastically increase the quality of numerical simulations of close highly radiative CWBs.
 Nevertheless, as seen from Fig. \ref{fig: prdcted_lum} and Table \ref{tab: windvel}, the {\it macroscopic} properties, and more specifically the intrinsic emission, of the interaction region in our simulation are in fair agreement with what is expected from purely theoretical considerations. This lends good support to our results, even if it is clear that a linear grid as the one we used is certainly not the most optimized approach to the current problem.\\

Other limitations are related to the 2-D nature of the simulations that inherently neglects the orbital motion of the binary. In a system such as \hda, the orbital velocities of a few hundred \kms\ compete with the pre-shock wind velocities that are only 2 to 4 times larger. Under such circumstances, the hypothesis of isotropic winds might also break down. Furthermore, in our simulations, the shell structure has radial progression velocity of only 100 \kms, much slower than the orbital motion. It is therefore very likely that the orbital motion might have a considerable influence on the geometry of the interaction region. One of the most obvious effects is the deflection of the interaction region. Due to the competition between the orbital motion and the radial velocities of the shell structure, we expect the interaction zone to be rather tilted. Another problem inherent to the 2-D nature of the simulation is that we let the flow evolve in a {\it frozen} orbital configuration. Therefore the X-ray emission obtained is definitely independent of the history of the system. The winds, escape flow and orbital motion of the \hda\ system seem to evolve on similar time-scales. In the real world, one can therefore reasonably expect that these phenomena compete or combine and that the resulting X-ray emission is correlated to the history of the different phenomena. We also note that tidal distortions of the stars in an eccentric binary may induce an asymmetric behaviour of their interaction with respect to periastron \citep{MK99}.\\

To conclude this section, the physics of the model itself can still be improved. Notably by including in the simulations other physical phenomena such as radiative braking \citep*{GOC97} and/or inhibition \citep{SP94}. 
In the particular case of the \hda\ binary, sudden radiative braking is unlikely to play a major role  as both stars of the system are of a broadly similar nature and therefore display similar luminosities and have winds of similar mass-loss rates and velocities. These winds will thus have similar radiative driving efficiencies (that is similar CAK type wind constants $k$ and $\alpha$). All these factors work against sudden radiative braking being important in this system and we can safely ignore it as a major effect for the studied object. 
A potentially more significant problem is the inhibition effect. Since the stars are quite close, the radiation of both stars will play an important role and could  alter the geometry of the wind-wind collision, both in terms of slowing and obliquely deflecting the colliding wind stream. Finally, the influence of the X-rays emitted by the collision on the winds of the stars and tidal distortions of the star surfaces that probably affect the structure of the winds are further effects that should also be addressed in CWB simulations. Properly accounting for all the above mentioned physical phenomena within the code is however far beyond our present purposes and  remains part of long term developments of CWB hydrodynamical models.\\

\section{Short term variability} \label{sect: st_var}
As suggested by the hydrodynamical simulations, the X-ray light curves of  \hda\ might display evidence of {\it short-term} variability, i.e.  variations on characteristic time scales ranging from a couple of minutes to a couple of hours and that are therefore much shorter than the orbital period. 
This section summarizes the results of the different methods implemented to search for short term variability within the collected \xmm\ data. For each pointing, we applied the following variability tests: i) we fitted polynomials of degree $n$, with $n=0$, 1 or 2, and we estimated both the quality $Q$ of the fit and the merit $F_\chi$ of including an additional term; ii) we  applied the Kolmogorov-Smirnov test to check the uniformity of the distribution function; iii) we used a corrected version of the \emph{probability of variability} test suggested by \citet[ see Appendix \ref{appendix}]{PZ02}; iv) finally, we also performed a \emph{Detrended Fluctuation Analysis} (DFA), a method that aims at searching for long-range correlation within time series. The latter three methods were applied on the series formed by photon arrival times. Though it was therefore not possible to account for a background correction, we restrain our analysis to PI events larger than 500 to minimize the impact of background events. Indeed, in the [500-10\,000] PI energy band, the background level only accounts for about 2\% of the detected counts and should therefore not be a critical issue for these tests.\\

\begin{figure}
\begin{center} 
\caption{Background corrected light curve (in \cnts) of the \hda\ system in the range PI $\in [500-10\,000]$ as presented in Fig. \ref{fig: glob_lc}. Detected local trends within the different sets of observation (see Table \ref{tab: chi2}) have been overplotted (plain lines). Dashed lines were computed assuming a 1-$\sigma$ difference from the best-fit slope values. The fitted local trends of  Obs. 2 and 6 (at respectively $\phi=0.66$ and $0.31$) are not significantly different from a constant level and are therefore not plotted on this figure.
\label{fig: trends}}
\end{center}
\end{figure}
 
\begin{table}
\centering
\caption{Best fit values of the slope whenever a linear model is the most appropriate to describe the light curve. The adopted bin size is 1000\,sec and the energy ranges approximately from 0.5 to 10\,keV (PI$\in$[500-10\,000]). All the values are expressed in $10^{-7}$ cnt\,s$^{-2}$.
\label{tab: chi2}}
\begin{tabular}{c c c c c}
\hline
\epic & Obs. 1 & Obs. 3 & Obs. 4 & Obs. 5 \\
\hline
\mos1 & $9.6\pm2.4$  & $-5.4\pm2.1$  & $-4.5\pm2.1$ & $4.7\pm2.2$ \\
\mos2 & $5.3\pm2.4$  & $-8.4\pm2.1$  & $-2.5\pm2.2$ & $7.6\pm2.2$ \\
\pn  & $11.0\pm4.0$  & $-19.5\pm3.6$ & $1.2\pm3.8$  & $11.6\pm3.7$ \\
\hline
\end{tabular}
\end{table}

\subsection{Polynomial fits} \label{ssect: chi}
For each pointing, we first extracted background corrected light curves from the filtered event list using time bin sizes ranging from 10 sec to 5000 sec in order to investigate the different time scales. We then fitted polynomials of degree $n=0$, 1 and 2 to the light curves. In doing so, we only consider here the detection of trends within a single 30\,ksec observation. The following criteria were used to estimate the adequation of the fitted model. First, we computed the merit $F_\chi$ of adding a supplementary term to a polynomial of degree $n$. This is given by : 
\begin{equation}
F_{\chi}=\frac{\chi^2_{n}-\chi^2_{n+1}}{\chi^2_{n+1}/(N-n-2)}
\end{equation}
where $N$ is the number of bins in the light curve \citep[see e.g.][]{Bevington}. As a ratio of two $\chi^2$ distributions, the statistic $F_\chi$ follows the Snedecor $F$-distribution with $\nu_1=1$ and $\nu_2=N-n-2$. We then required that $F_\chi$ exceeds a threshold value $F_\chi \ge f$, corresponding to the probability $P_F(f;\nu_1,\nu_2)=0.95$. Second, the model should reasonably fit the light curve, i.e. the quality $Q$ of the fit -- which is given by:
\begin{equation}
Q(\frac{\nu}{2},\frac{\chi^2}{2})=1-P(\frac{\nu}{2},\frac{\chi^2}{2})
\end{equation}
where $P$ is the incomplete gamma function and $\nu$, the number of degrees of freedom --  should yield a reasonable value \citep[see e.g.][]{Press}. Table \ref{tab: chi2} reports the values of the trends whenever a linear model was the most appropriate to describe the data. Though these values seem to be quite small in terms of count rate, they might account for a 10 to 20 per cent variation over a 30\,ksec exposure and are therefore significant. The inclusion of an additional square term turns out to be unnecessary in most of the cases and only marginally improves the quality of the fit in the very few cases where an  additional term could have been relevant. 
These constraints on the values of the first order derivative of the flux might be useful in testing more evolved models where, for example, the orbital evolution of the system with time is accounted for.\\

We note that the slopes observed for the two \mos\ instruments are in reasonable agreement, whereas the \pn\ best fit slope is usually steeper. This might be explained by the different response of the \mos\ and \pn\ and, indeed, the slopes observed in the S and M energy bands are often significantly different. This is further in agreement with the fact that the global variation of the flux goes along with a modification of the spectral properties of \hda\ (see Sect \ref{ssect: fits}).
Figure \ref{fig: trends} displays the background corrected lightcurve of Fig. \ref{fig: glob_lc} on which the detected local trends have been represented. It is clear from this figure that most of the detected trends are indeed in line with expectations from the longer timescale orbital variation, but for Obs. 5 at $\phi=0.15$. Indeed the steep trend detected in Obs. 5 count rates seems to initiate a rising of the flux towards apastron, as it is seen in the model predictions. However, Obs. 6 at $\phi=0.31$ does clearly not follow that trend and shows a further decrease of the flux. The exact reason of the limitation of the flux increase towards apastron remains at this stage largely unexplained. However the simulations of the previous section indicate that part of the answer might be linked to the orbital motion of the system.\\

\subsection{Kolmogorov-Smirnov} \label{ssect: KS}
To check for constancy, the Kolmogorov-Smirnov test was applied on the series formed by the time difference between two photon arrival. We accounted for the Bad Time Intervals (BTIs) within a single pointing by subtracting the respective BTI durations to the arrival time of the photons that were chronologically following the BTIs. This is equivalent to assuming that the detector is switched off during the BTIs. Such a procedure might actually favour the rejection of the null hypothesis of a uniform distribution if and only if the underlying distribution is not uniform. In this sense, we might therefore consider that the adopted elimination of the BTIs does not bias the  Kolmogorov-Smirnov test.\\

The Kolmogorov-Smirnov test reveals significant variability ($>95$ per cent) for those pointings that display trends, as shown in the previous paragraph (Sect. \ref{ssect: chi}). Inspection of the time series further allows to assert that the Kolmogorov-Smirnov test actually detects the local trends within the relevant pointings, in agreement with the results of the polynomial fits (Sect. \ref{ssect: chi}).\\

\subsection{Probability of Variability} \label{ssect: pov}
We applied a modified version of the \citet{PZ02} probability of variability test (see Appendix \ref{appendix}) to the \hda\ and background event lists for which the PI events are in the range [500-10\,000[. One of the advantages of this method is to directly account for the poissonian distribution of low count rate observations and it is therefore particularly designed to investigate any background variability. As the values of the adopted binning depends upon the expected number of counts $<N>$, we choose $<N>$ to range from two to a few thousands in order to investigate different time scales. Indeed a larger value for $<N>$, and hence for the time bin size, favours the detection of long range variation. On the other hand, small  values of $<N>$ will tend to enhance small scale variations that could be diluted or averaged out in the first case. The choice for $<N>$ could thus enhance or hide variability within a particular characteristic time scale. We adopt a confidence level of $pov>0.99$ for the rejection of the null hypothesis of a uniform distribution of the counts throughout the exposure. \\

Concerning the \hda\ binary system, clear cut positive results occur for Obs. 3  ($\phi=0.81$) for the \epicmos2 and \pn\ instruments while variability in the  \mos1 is only marginally detected ($0.90<pov<0.99$). The variability seems to occur on time scales above 1000\,sec and might be related to the detection of a `steep' trend in the related data. The background of Obs. 4  ($\phi=0.00$) is also variable, especially for the \epicpn\ instrument though the corresponding \hda\ event list for Obs. 4 gives a negative result under the $pov$ test. Finally marginal variability ($0.90<pov<0.99$) is also detected for Obs. 1  ($\phi=0.54$) and might again be related to the detected trend. \\

In conclusion, the three methods described in Sections \ref{ssect: chi} to \ref{ssect: pov} give fully consistent results and prove that local trends, presumably linked to the orbital variability, are clearly present within four out of the six pointings. However, shorter time scale variability such as those seen in the results of the models described in Section \ref{sect: simul} has not been found. We might wonder if the local trends do not alter the properties of the time series and therefore obscure the possible intrinsic variability of the wind-wind collision. One of the recently developed methods that allows to characterise a signal intrinsic variability regardless of possible local trends is the  {\it Detrended Fluctuation Analysis}. This will be the focus of the next section.\\

\begin{figure}
\begin{center} 
\caption{
Integrated time series $y(i)=\sum_{j=1}^{i}\left( \Delta t_j -\overline{\Delta t} \right)$ built from the \epicpn\ observation at $\phi=0.81$ in two energy ranges. R1: PI$\in[500-10\,000]$ (solid line) and R3: PI$\in[2000-10\,000]$ (dotted line). Left and bottom scales refer to R1 while right and top scales refer to R3.
\label{fig: tserie}}
\end{center}
\end{figure}

\begin{figure}
\begin{center} 
\caption{Evolution of $F(k)$ with the window size $k$ plotted on a $\log-\log$ scale diagram. The data set is built from \epicpn\ events at $\phi=0.81$. Different symbols refer to different energy ranges. Notations R1 to R3 have the same meaning as in Fig. \ref{fig: tserie} and in Table \ref{tab: dfa_table}.
\label{fig: dfa_plot}}
\end{center}
\end{figure}

\subsection{Detrended Fluctuation Analysis} \label{ssect: dfa}
The {\it Detrended Fluctuation Analysis} (DFA) is a method that aims at searching for long-range correlations in noisy signals \citep{dfa1,dfa2} and is based on the variance analysis of the fluctuations around the local trend. Assuming a self-similar sequence $\{y(i) \}$ of $N$ elements, we divide the series into $N_\mathrm{w}$ non-overlapping windows of $k=N/N_\mathrm{w}$ elements. We then compute the mean square fluctuation around the local trend $y_l$ in the considered window : 
\begin{equation}
F^2_k(l)=\frac{1}{k}\sum_{i=l k+1}^{(l+1) k}\left(y(i)-y_l(i)\right)^2
\label{eq: dfra_eq1}
\end{equation}
where $y_l$ is the least-square linear fit function in the window $l$ ($l=0,1,2, ..., N_\mathrm{w}-1$). We then average $F^2_k(l)$ over the $N_\mathrm{w}$ windows of size $k$. Repeating the operation for different $k$ values yields a power-law relation between the root mean square fluctuation function $F$  and the window size $k$ \citep[see e.g.][]{KKRHB01}:
\begin{equation}
F(k)=\sqrt{<F^2_k>}\sim k^{\alpha}
\label{eq: dfa_eq2}
\end{equation}
where $\alpha$ is the Hurst exponent, also called the long-range correlation exponent, which is typical of the series behaviour. For a purely random walk (e.g. Brownian motion), $\alpha = \frac{1}{2}$. For $\alpha < \frac{1}{2}$, the signal harbours \emph{anti-persistent} correlations while for  $\alpha > \frac{1}{2}$, \emph{persistent} correlations are present \citep[see e.g.][]{Feder}.
This method has proved its worth in a wide variety of domains such as physiology \citep{Ashkenazy}, biophysics \citep{Stanley} and econophysics \citep{vdw}. \citet{Stanley} also compared its performances to other more widespread techniques (correlation function, power spectrum, rescaled range (R/S) analysis) and report that the performances of the DFA method merely rest on its increased ability to reduce the noise in the determination of the correlation exponent $\alpha$ and to the fact that the method is mostly insensitive to both local and global trends. 
Some authors \citep[e.g.][]{HICCS01} suggest the use of polynomials of degree $n$ (DFA-$n$ analysis), with $n$ larger than one, instead of the linear local trend $y_l$. We however already showed (Sect. \ref{ssect: chi}) that linear models are the most successful in describing the trend within a single pointing. We therefore limit our DFA analysis to the so-called DFA-1 approach. \\

\begin{table}
\centering
\caption{
Values of the Hurst exponent (or long-range correlation exponent) $\alpha$ as obtained from a least-square linear fit on a $\log-\log$ scale for the three \epic\ instruments and for different energy ranges. R1: PI$\in [500-10\,000]$;  R2: PI$\in [1000-10\,000]$;  R3: PI$\in [2000-10\,000]$.
\label{tab: dfa_table}
}
\begin{tabular}{c c c c c}
\hline
Obs. \# & PI & \mos1 &\mos2 & \pn \\
\hline
1 & R1 & 0.540$\pm$0.006 & 0.540$\pm$0.004 &0.512$\pm$0.004 \\
  & R2 & 0.532$\pm$0.009 & 0.547$\pm$0.006 &0.519$\pm$0.006 \\
  & R3 & 0.555$\pm$0.016 & 0.535$\pm$0.016 &0.537$\pm$0.008 \\
\hline
2 & R1 & 0.496$\pm$0.008 & 0.495$\pm$0.009 &0.506$\pm$0.006 \\
  & R2 & 0.494$\pm$0.008 & 0.510$\pm$0.007 &0.515$\pm$0.006 \\
  & R3 & 0.470$\pm$0.021 & 0.681$\pm$0.056 &0.674$\pm$0.019 \\
\hline
3 & R1 & 0.502$\pm$0.009 & 0.500$\pm$0.007 &0.514$\pm$0.007 \\
  & R2 & 0.479$\pm$0.014 & 0.518$\pm$0.012 &0.503$\pm$0.006 \\
  & R3 & 0.563$\pm$0.021 & 0.589$\pm$0.025 &0.568$\pm$0.014 \\
\hline
4 & R1 & 0.521$\pm$0.009 & 0.496$\pm$0.009 &0.546$\pm$0.005 \\
  & R2 & 0.496$\pm$0.011 & 0.503$\pm$0.012 &0.514$\pm$0.013 \\
  & R3 & 0.549$\pm$0.021 & 0.583$\pm$0.025 &0.517$\pm$0.022 \\
\hline
5 & R1 & 0.553$\pm$0.011 & 0.521$\pm$0.009 &0.502$\pm$0.004 \\
  & R2 & 0.497$\pm$0.011 & 0.495$\pm$0.011 &0.461$\pm$0.007 \\
  & R3 & 0.535$\pm$0.021 & 0.687$\pm$0.027 &0.496$\pm$0.017 \\
\hline
6 & R1 & 0.548$\pm$0.007 & 0.477$\pm$0.008 &0.500$\pm$0.005 \\
  & R2 & 0.521$\pm$0.013 & 0.475$\pm$0.010 &0.508$\pm$0.007 \\
  & R3 & 0.547$\pm$0.019 & 0.494$\pm$0.022 &0.514$\pm$0.015 \\
\hline
\end{tabular}
\end{table}

As the extension of the method over several decades of the window size $k$ is crucial to properly estimate the correlation exponent, we construct our time series using the time intervals $\Delta t_i$ that separate the arrivals of two successive photons in the adopted region for \hda\ (see Sect. \ref{ssect: epic_data_red}). The constructed time series are then mapped onto a self-similar process by integration:
\begin{equation}
y(i)=\sum_{j=1}^{i} \left( \Delta t_j - \overline{\Delta t} \right)
\end{equation}
where $\overline{\Delta t}$ is the average of the time intervals $\Delta t$ over the whole time series. Figure \ref{fig: tserie} shows an example of such a series corresponding to the \epicpn\ observation of \hda\ at $\phi=0.81$.
For a window size ranging approximately from $k=4$ to $k=N/4$, we then compute the average $F(k)$ values.
 As can be seen from Fig. \ref{fig: dfa_plot}, this yields a power-law relation $F(k)\sim k^{\alpha}$. We then adjust a linear relation on a $\log-\log$ scale to estimate the value of the power-law parameter $\alpha$.
Figure \ref{fig: dfa_plot} presents a typical evolution of $F(k)$ with the number $k$ of elements in a window and Table \ref{tab: dfa_table} gives the values of the Hurst exponent $\alpha$ as obtained from a least-square linear fit. We performed a similar analysis in different energy bands. While in the range PI $\in [500-10\,000]$, $\alpha$ takes values around 0.5,  it appears that the self-similarity parameter $\alpha$ tends to reach somewhat higher values when restricting the data set to higher energies. This suggests that persistent long-range correlation might be present in the higher energy tail of the flux. As indicated by our simulations, this hard tail above 2\,keV might find its origin in the interaction region. \\

\citet{SM86} showed that several facets of turbulence could be related to fractals. In particular, \citet{MS87} reported that energy dissipation in turbulent processes  displays long-range correlations with $\alpha < 0.5$. Their best model yields $\alpha \approx 0.45$. The apparent increase of the self-similarity coefficient with the energy in our \xmm\ time series tends to yield $\alpha$ values slightly larger than 0.5 and might therefore not be directly related to turbulence occurring within the wind-wind interaction region. 
Though the results of the DFA method presented in this paper are not clear cut, they however constitute an encouragement to apply this kind of technique to a wide variety of astrophysical time series analyses.\\

\section{Summary and conclusions}
\label{sect: concl}
In this paper, we presented the results of an \xmmnew\ campaign on the massive CWB binary \hda\ that was performed as part of the GT observation time of the OM-consortium. The campaign was split into six separate pointings, for a total duration of 180\,ksec and the phase resolution of the observations is probably one of the best ever performed on an O+O colliding wind binary. We showed that the X-ray flux emitted by \hda\ presents a clear enhancement of about 50 per cent culminating slightly after apastron passage and that the relative amplitude of this increase is also dependent on the considered energy range. Comparison with previous \rosatpspc\ data provides a good support for a phase-locked behaviour of the observed modulations that further remains stable on time scales of at least 10 years. The \xmmepic\ spectra are relatively soft and present their maximum peak between  0.8 and 0.9\,keV. They are reasonably well described by a two-temperature \mekal\ model in the range [0.5--2.5\,keV]. The temperature of the low-energy component is located between 0.26 and 0.31\,keV while the second component has a temperature ranging from 0.62 to 0.71\,keV (see Table \ref{tab: 2t}). A third component of higher energy is clearly seen in several spectra, though its nature could not be unambiguously established. It could indeed correspond to a power-law tail characterized by a photon index $\Gamma\sim3.7$ or, though with a slightly lower confidence, to a thermal component with a temperature of a few keV. If the existence of the non-thermal component is indeed confirmed, then its spectral index is much larger than the 1.5 value expected for a strong shock. Indeed, \citet{CW91} expect such a component to be produced by inverse Compton scattering radiation emitted by a population of relativistic electrons that would have been accelerated in shocks through the first-order Fermi process. This result is similar to what has been observed for other early type O stars such as \sgr\ and \hdb\ \citep{Rauw02_9sgr,Rauw02} and the problem definitely deserves further attention. Combining the data from all six pointings however reveals that the Fe line at 6.7\,keV is most probably present. This implies that at least part of the high energy tail is likely to have a thermal origin.\\

We also performed numerical hydrodynamical simulations that closely reproduce the geometrical and orbital configuration of the \hda\ system at the six phases corresponding to the different \xmm\ pointings. The resulting hydrodynamical maps reveal that the interaction region is highly unstable and, as we progress away from the binary axis, it develops a filamentary structure that alternates thin but dense shells with larger low density regions. The computed averaged intrinsic flux produced by such a configuration is in excellent agreement with the analytical estimate for the X-ray emission resulting from the interaction region of two identical winds. For a given configuration, the absorbed X-ray luminosity predicted by the model further present clear variations on time scales of a couple of hours. Finally, the predicted luminosities show a clear enhancement by about one order of magnitude around apastron. Adding the expected intrinsic contribution of the two O-type stars of the system, the total X-ray luminosity reasonably well reproduces both the observed luminosities and the amplitude and phase of their variation. Due to the 2-D nature of the simulations and the fact that we neglect the orbital motion, the predicted light curve is symmetric around apastron and could therefore not properly render the observed decrease of the flux beyond periastron phase. Part of the answer to this problem may probably be found in the deflection of the interaction region due to the orbital motion and subsequent absorption.\\

The energy distribution predicted by the hydrodynamical simulations are peaked around 3\,keV and are therefore much harder than the observed \epic\ spectra. The models under-predict the flux at low-energy, which is most probably due to the fact that the X-ray contribution of the two O-stars has been neglected. Similarly, the hard-energy tail seems to be overestimated in the model prediction, though the observed energy distribution still remains within the boundaries of the model.
We are conscious that our simulations rest on several oversimplified hypotheses such as constant wind velocities and \emph{frozen} orbital configurations, whose effects may influence the output of our models. We are however confident that our results provide a good approximation to the `reality'. Another limitation that we are facing is the extremely thin interaction region that could not be resolved with the adopted numerical mesh. One significant improvement for hydrodynamical simulations of highly radiative colliding wind interactions could therefore be the use of adaptative meshes.\\

Using different techniques, we also investigated the X-ray variability of the source on short-time scales. Linear trends are definitely present in several data sets and helped to constrain the first order derivative of the flux. We also applied the  \emph{Detrended Fluctuation Analysis}, a method that has proved its merits in other domains of sciences. Its application to the \epic\ data suggests that long-range correlations might be present in the series formed by the photons arrival times, especially while restricting the analysis to the hard energy tail of the data. This however does not seem to be directly related to  possible turbulent processes occurring within the interaction region. Short-time variability with time scales of a couple of hours as seen in the predictions of the simulations could finally not be brought into light. On one hand, it is possible that these variations are diluted by the additional flux from the two stars or perturbed by some physical phenomenon not accounted for in the simulations, such as the orbital motion. On the other hand, the predicted variability might be an artefact resulting from the 2-D nature of the simulations in which the symmetry might lead to an overestimate of the variability. Again 3-D simulations are required to clarify this question.\\

Finally, we emphasize that this campaign on the \hda\ massive binary is one of the most complete phase-resolved X-ray studies of a colliding wind O+O binary ever performed. It further outlines the insight in the physics of the phenomenon that can be gained from a detailed comparison of the observations with hydrodynamical models. It also emphasizes that more detailed hydrodynamical models, possibly using 3D configuration and accounting for the orbital motion, are needed to achieve a better understanding of the local physics that governs the X-ray emission of colliding wind binaries. The need to extend this kind of study to a growing number of CWB systems, either O+O or WR+O binaries, is obvious to achieve a detailed understanding both of the hot star winds and of their interaction within CWB systems.\\

\section*{Acknowledgments}
The authors wish to thank the referee, Dr. Ken Gayley, for careful reading of the manuscript. HS would like to express his thanks to David B. Henley for very helpful discussions on the hydrodynamical simulations. He is also grateful towards Dr. Nicolas Vandewalle for his assistance in handling the DFA method and to Micha\"el De Becker and Ya\"el Naz\'e for comments on the variability analysis techniques. The Li\`ege team acknowledges support from the Fonds National de la Recherche Scientifique (Belgium) and through the {\sc PRODEX XMM-OM} and Integral projects. This research is also partly supported by contracts P4/05 and P5/36 ``P\^ole d'Attraction Interuniversitaire'' (Belgian Federal Science Policy Office). IRS acknowledges funding from PPARC. HS's thanks also go to the Communaut\'e Fran\c{c}aise de Belgique for travel support (``Bourse de Voyage''), to PPARC for funding a visit to Birmingham and to the University of Li\`ege for taking care of his integration and for {\it generously} providing heat and electricity.\\

\appendix
\section{The \emph{probability of variability} test revisited} \label{appendix}
\subsection{\citeauthor{PZ02} $pov$ test}
The \emph{probability of variability} test suggested by \citet{PZ02} basically gives the confidence level -- called the \emph{probability of variability} ($pov$) -- to reject the null hypothesis of a constant count rate for the considered source. To do so, given the mean count rate, they first estimate the mean time interval $\Delta t$ during which two photons are expected to be detected by the instrument. Then, using a sliding window, they determine the maximum number of counts  $N_\mathrm{max}$ received during such a period $\Delta t$. Finally, they estimate the probability $p$ that a poissonian distribution of mean $2$ gives at least $N_\mathrm{max}$ counts in a $\Delta t$ interval. This is simply given by the formula :
\begin{equation}
p=1-q = 1 - \sum_{k=0}^{N_\mathrm{max}-1} e^{-2} \frac{2^k}{k!}
\end{equation}
The \emph{probability of variability} given by $pov=1-p$ estimates whether such a high number of counts $N_\mathrm{max}$ as observed is rather improbable, which is expressed by a large value of the $pov$ and suggests a significant variability in the data. \\

However, in their method, \citeauthor{PZ02} do not account for the fact that, actually, they are not performing a single experiment (i.e. observing a single interval) but rather $n_\mathrm{int}$  of them, where the value for $n_\mathrm{int}$  can be quite large (a few thousands in the case of our present observations). Indeed, repeating a counting experiment increases the probability to observe a number of counts that seems to  significantly deviate from the expected values but still is in fair agreement with the statistics theory. \citeauthor{PZ02} also failed to consider a more subtle aspect: the fact that their ``sliding window'' approach implies that the observed intervals are overlapping and thus that the counting of photons in the different windows are not independent. In its current form, the $pov$ confidence level used by \citeauthor{PZ02} is biased towards larger values (see Table \ref{tab: pov_expl} for an example) and clearly favours the rejection of the null hypothesis of non variability, which might lead to a dramatic increase of type I error occurrences. \\

\begin{table}
\centering
\caption{Example of the dependence of the $pov$ estimator with the number of  observed intervals $n_\mathrm{int}$. First column gives the values of $n_\mathrm{int}$. Second to fourth columns provide  the $pov$ confidence level for a maximum observed counts $N_\mathrm{max}$ respectively of 5, 6 and 8 as computed following our revised version with $<N>=2$ (see Sect. \ref{ssect: pov_hs}). \citeauthor{PZ02} would have obtained values corresponding to $n_\mathrm{int}=1$ for each case. 
\label{tab: pov_expl}
}
\begin{tabular}{c c c c}
\hline
$n_\mathrm{int}$  & \multicolumn{3}{c}{\emph{probability of variability}} \\
           & $N_\mathrm{max}=5$  & $N_\mathrm{max}=6$  & $N_\mathrm{max}=8$ \\
\hline
   1 & 0.9473 & 0.9834 & 0.9989 \\
\hline
   2 & 0.8975 & 0.9671 & 0.9978 \\
   5 & 0.7630 & 0.9197 & 0.9945 \\
  10 & 0.5822 & 0.8458 & 0.9891 \\
  20 & 0.3390 & 0.7154 & 0.9783 \\
 100 & 0.0045 & 0.1882 & 0.8961 \\
1000 & 0.0000 & 0.0000 & 0.3338 \\
\hline
\end{tabular}
\end{table}

\subsection{The $pov$ revisited}\label{ssect: pov_hs} 
Let $\Delta t$ be the time period during which the expected number of counts is $<N>$ ($<N>=2$ in \citeauthor{PZ02} original method). Let  $n_\mathrm{int}$ be the number of non overlapping intervals of size $\Delta t$ among which we distributed our data. Finally, let $N_\mathrm{max}$ be the maximum number of counts found in a single interval among the $n_\mathrm{int}$ ones. Therefore, as the $n_\mathrm{int}$ countings might reasonably be considered as independent and neglecting the fact that $\Delta t$  has been estimated from the time series, the probability to find by chance  (i.e. due to statistical fluctuations) {\it at least} $N_\mathrm{max}$ counts in {\it one or more} intervals among the $n_\mathrm{int}$ observed ones is now given by:
\begin{equation}
P=1-q^{n_\mathrm{int}}=1 - \left(\sum_{k=0}^{N_\mathrm{max}-1} e^{-<N>} \frac{<N>^k}{k!} \right)^{n_\mathrm{int}}
\end{equation}
while the \emph{probability of variability}, i.e. the probability that such a high number of counts $N_\mathrm{max}$ is not due to statistical fluctuations, is given by $pov = 1-P$. As of course $q<1$, $P$ increases with the number of intervals $n_\mathrm{int}$ and the $pov$ might thus be drastically reduced (see Table \ref{tab: pov_expl}).  \\

In the regard of the present considerations, the $pov$ levels found by \citet{PZ02} in their Section 7 on the variability of the studied objects are therefore overestimated or, at least, could not be compared to a unique rejection criterion as larger values of their $pov$ confidence level are naturally expected for data displaying a larger number of counts. This is especially true concerning the objects with the higher number of counts (N-045-02 and L-312) for which the large values of their $pov$ confidence level found ($pov=0.95$) are most probably due to statistical fluctuations rather than reflecting any intrinsic variability of the sources.\\

\label{lastpage}
\end{document}